\documentclass[%
preprint,
amsmath,amssymb,
aps,
prfluids,
]{revtex4-2}

\usepackage{graphicx}% Include figure files
\usepackage{dcolumn}% Align table columns on decimal point
\usepackage{bm}% bold math
\usepackage{gensymb}
\usepackage{textcomp}
\usepackage{todonotes}
\usepackage{physics}
\usepackage[colorlinks=true,citecolor=blue,urlcolor=blue,linkcolor=blue]{hyperref}

\usepackage[normalem]{ulem}

\usepackage{cancel}
\usepackage{subeqnarray}
\usepackage{mathtools}
\def\be{\begin{equation}}
\def\ee{\end{equation}}

\usepackage{verbatim}

\usepackage{cleveref}

\crefformat{section}{\S#2#1#3} % see manual of cleveref, section 8.2.1
\crefformat{subsection}{\S#2#1#3}
\crefformat{subsubsection}{\S#2#1#3}

\def\ee{{\rm e}}

\begin{document}

\preprint{APS/123-QED}

\title{5-wave interactions in inertia-gravity waves}
% Force line breaks with \\

\author{Saranraj Gururaj}%
\email{gmsaranraj@gmail.com}
\affiliation{School of Science and Engineering, University of Dundee, Dundee DD1 4HN, UK.}
\author{Anirban Guha}%
\affiliation{School of Science and Engineering, University of Dundee, Dundee DD1 4HN, UK.}


%\affiliation{
%\affiliation{
%\affiliation{%


\date{\today}% It is always \today, today,

\begin{abstract}

 In oceans, multiple energetic inertia-gravity waves often coexist {in a region}.  
 % For example, leftward and rightward propagating inertia-gravity waves generated by tide-topography interactions overlap in the vicinity of the topography. 
 In this paper, we study the stability of two coexisting plane inertia-gravity waves (hereafter, primary waves), with the same frequencies ($\omega_1$) and wavevector norms, in a region of constant background stratification {(denoted by $N$)}. Specifically, we explore the decay of two primary waves through triadic resonant instabilities (TRIs) in cases where the primary waves do not resonantly interact with each other. 
  % When two coexisting primary waves undergo triadic resonant instability, they both can force the same secondary wave, and this results in a 5-wave system (5WS) composed of two different triads. Each of these triads consists of a primary wave and two secondary waves, with one secondary wave common between the two triads. 
{Two coexisting primary waves undergoing triadic resonant instability can force two secondary waves each, and this results in two 3-wave systems (3WS). In some cases, two primary waves can have a common secondary wave, and this results in a 5-wave system (5WS) composed of two different triads.} 
 We show that 5WSs are the dominant instabilities with higher growth rates than standard triads for a wide range of Coriolis frequency values ($f$). For 2D cases, 5WSs have higher growth rates than triads for $f/\omega_1\gtrapprox0.3$ and for primary waves with the same horizontal (vertical) wavenumber but with {opposite} vertical (horizontal) wavenumber. {Similar results are observed for 3D cases where the primary waves are not on the same vertical plane}. Numerical simulations match the theoretical growth rates of 5WSs for a wide range of latitudes, except when $f/\omega_1\approx0.5$ (critical latitude). Using theory and simulations, we show that the maximum growth rate near the critical latitude is approximately twice the maximum growth rate of all triads.

 %%%%%%%%%%%%%%%%%%%%%%%%%%%%%%%%%%%%%%%%%%%%%%%%%%%%%%%%%%%%%%%%%%%%%%%%%
 
\end{abstract}

\maketitle

\section{Introduction}\label{section:1}

  { Inertia-gravity waves (hereafter simply referred to as internal waves) play a major role in sustaining the Meridional Overturning Circulation {through} diapycnal mixing \citep{MUNK1998,ferrari_2008}.} Wave-wave interaction is estimated to be one of the most dominant mechanisms through which internal waves' energy cascades to small length scales \citep{DE_2019}, where it can cause {irreversible} mixing. As a result, understanding wave-wave interactions {is} important to model the internal waves' energy cascade.

The stability of a plane internal wave with small steepness has been studied extensively. Here, the steepness is defined as the ratio of the maximum wave-induced velocity along the horizontal direction to the phase velocity of the wave along the same direction \citep{koudella_staquet_2006}. Steepness is used as a non-dimensional measure of the wave's amplitude. A primary internal wave with small steepness is unstable to secondary waves through triad interactions if the secondary waves' frequencies are lesser than the primary wave's frequency \citep{hasselmann_1967}. Moreover, the three waves' frequencies and wavevectors should also satisfy the resonant triad conditions: $\mathbf{k}_{1} = \pm \mathbf{k}_{2} \pm \mathbf{k}_{3}$ and $\omega_1 = \pm \omega_2 \pm \omega_3$ \citep{thorpe_1966,davis_acrivos_1967,hasselmann_1967},  where the primary and secondary waves are denoted by subscripts 1, and 2
and 3, respectively. 

For primary waves with small steepness, a 2D stability analysis is sufficient to find the most dominant instability \citep{klostermeyer}. The most unstable secondary wave combination depends on the ratio of primary wave frequency to the  {buoyancy frequency $N$ \citep{Sonmor_1997}, kinematic viscosity $\nu$ \citep{bourget,bourget_width_2014}, and Coriolis frequency $f$} \citep{young,maurer_joubaud_odier_2016}.   {In a non-rotating, inviscid fluid}, the wavevectors of the most unstable secondary wave combination satisfy $|\mathbf{k}_{3}| < |\mathbf{k}_{1}| < |\mathbf{k}_{2}|$ for $\omega_1/N < 0.68$ \citep{Sonmor_1997}.   However, the wavevectors of the most unstable secondary waves satisfy $ |\mathbf{k}_{2}| \approx |\mathbf{k}_{3}| \gg |\mathbf{k}_{1}|$ for $\omega_1/N > 0.68$ under inviscid conditions.  This instability is called Parametric Subharmonic Instability (PSI) \citep{Mackinnon_2005,young}. PSI is a special type of triad interaction where $\omega_2 \approx \omega_3 \approx \omega_1/2$. In the oceans, viscosity has very little impact on large length scale waves. In contrast, viscosity can play an important role in deciding the specific combination of waves that drain energy from the primary wave under laboratory conditions \citep{bourget}.  {Viscosity is one of the main reasons PSI is not observed in laboratory experiments \citep{bourget}.}

For internal wave triads, rotational effects can be very important for a wide range of latitudes, especially near the critical latitude (where $f \approx \omega_1/2$). Near the critical latitude, for any $\omega_1/N$, the primary wave gives its energy to waves whose frequency is close to the inertial frequency. Moreover, the near-inertial waves have very small vertical length scales that can lead to increased kinetic energy dissipation \citep{richet_2018}. Semidiurnal mode-1 internal waves have been observed to lose a non-negligible portion of their energy as they pass through the critical latitude \citep{Mackinnon_2005,Alford_2007,Haze_2011}. Moreover, when semidiurnal internal wave modes interact with an ambient wave field that follows the Garrett-Munk spectrum, their decay is fastest near the critical latitude \citep{hibiya_1998,decay_onuki,olbers_2020}.

In this paper, we study the stability of two weakly nonlinear plane primary waves that coexist in a region. The motivation for this study stems from the fact that primary internal waves generated in different locations often meet/overlap in the oceans. For example, tide-topography interactions result in the generation of internal waves that propagate in horizontally opposite directions, and these waves overlap/coexist above the topography, c.f. \citet[figure 7]{sonya_nik_2011}.

\begin{figure}
 \centering{\includegraphics[width=1.0\textwidth]{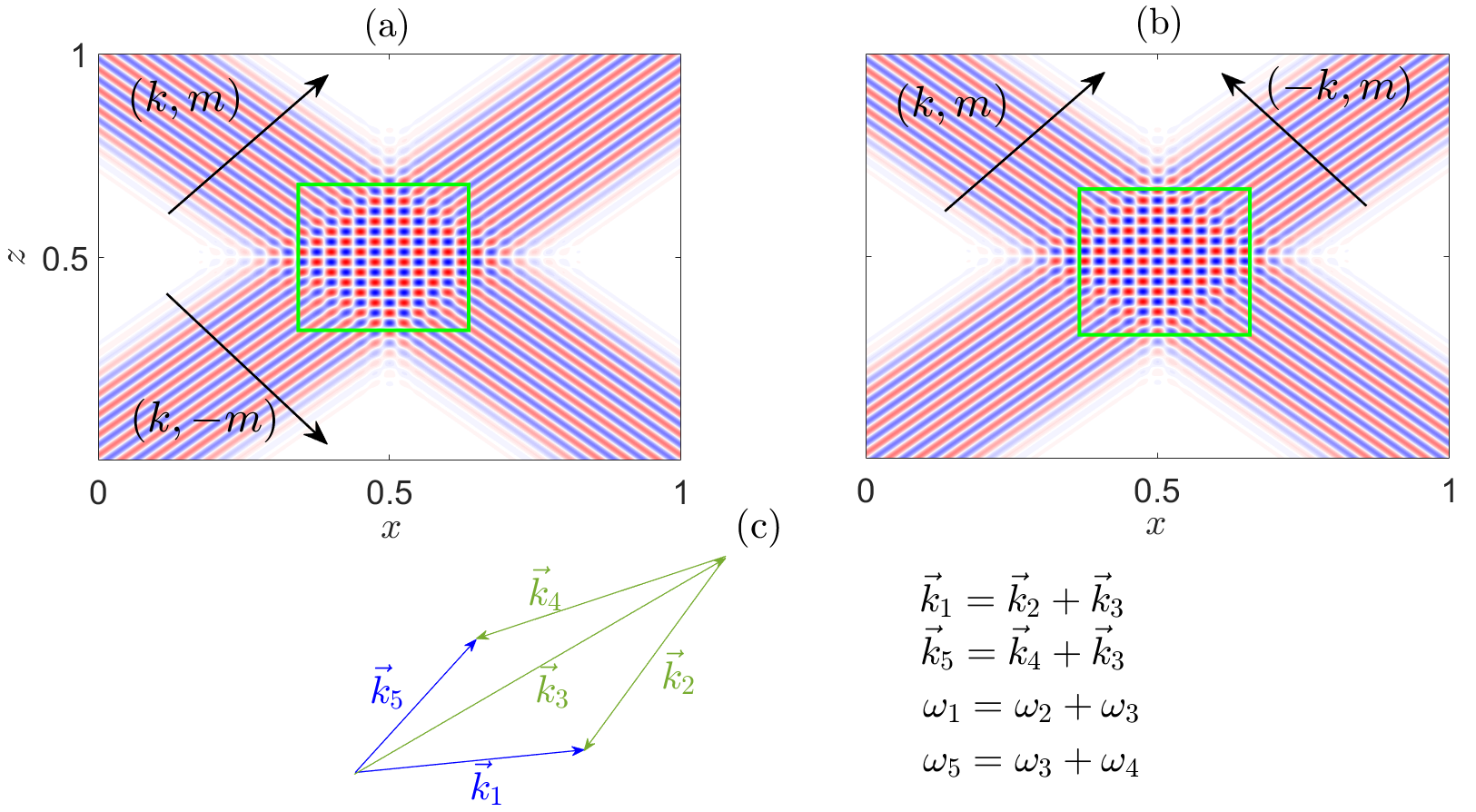}}
 \caption{ Examples of different orientations of propagating primary waves: in (a) vertically 
 and (b) horizontally opposite directions, with the intersection region marked in green.  (c) Frequency and wavevector triad conditions that are satisfied between the 5 waves that are involved in the interaction region. Waves 1 and 5 are primary waves, while waves 2,3, and 4 are secondary waves, with wave-3 being the common secondary wave. }
  \label{fig:5_wave_triad_schematic}
\end{figure}

When two energetic primary waves meet in a region, they can resonantly interact with each other. The collision of internal wave beams is an example of direct interaction between the primary waves and has been studied  extensively\citep{tabaei_akylas_lamb_2005,Jiang_Marcus_2009,akylas_karimi_2012}. Experimental studies on internal wave modes also show that two low modes with the same frequency can interact and produce a third wave \citep{Hus_2020}. 
 {Primary waves, however, do not always resonantly interact with each other and form a triad. In such cases, each primary wave would still be susceptible to triad interactions leading to the growth of secondary waves, and this is the setting explored in this paper.} Specifically, we focus on the 5-wave system instability. In this instability, five waves (two primary and three secondary waves) are involved, and two distinct triads are formed between the five waves.  Note that this implies that one secondary wave is part of two different triads and is forced by both primary waves.  {Some examples of overlapping primary waves are given in figures \ref{fig:5_wave_triad_schematic}(a)--\ref{fig:5_wave_triad_schematic}(b). The scenarios shown in figure 1 can easily occur in the oceans when internal waves are generated by tide-topography interactions. In both figures, the region enclosed by the green box is a potential location for a 5-wave system. The wavevector and frequency conditions satisfied in a 5-wave system are given in figure \ref{fig:5_wave_triad_schematic}(c).}   {Note that the two primary waves, in the weakly nonlinear limit, can always have two independent 3-wave interactions. In other words, one primary wave can interact with its own two secondary waves without the influence of the other primary wave. Hence, we aim to find which type of instability (5-wave or 3-wave systems) is the most dominant.}  {In this paper, we only focus on plane waves, and we do not consider finite width beams. An interaction involving two primary beams is a more complicated scenario because it involves two different spectra of waves. For simplicity, we chose to begin with plane waves.}   {For two primary plane waves, there are only two wavevectors, and as a result, the problem can be explored in depth. Moreover, the results should be applicable for wide internal wave beams since wide beams essentially behave like plane waves.}

In the context of internal waves,  5-wave systems have been studied recently \cite{common_parent_wave,5_wave_common_daughter_wave}. \citet{common_parent_wave} focus on 5-wave systems where the same primary wave generates four different secondary waves, which is not the focus of this paper. \citet{5_wave_common_daughter_wave} explore 5-wave systems that consist of two primary waves and three secondary waves, but their focus is on rogue wave generation. They study the 5-wave systems in a 2D setting without the rotational effects. Moreover, no detailed study was conducted on the growth rates. In this paper, we consider a 3D setting with rotational effects, which is observed to be important in our case. The primary focus is on the growth rates of the secondary waves and to understand scenarios in which the 5-wave system instability is faster than the 3-wave system instability (standard triads). In our study, the frequencies of the two primary waves are always assumed to be the same {(i.e. $\omega_1 = \omega_5$)}, and this assumption can be important in an oceanographic context since internal waves generated by the same tide have the same frequency. The paper is organized as follows. In \S \ref{Section:2}, we use multiple scale analysis to simplify the 3D, Boussinesq Navier-Stokes equations in the $f-$plane and derive the wave amplitude equations. Expressions for growth rates are provided.  In \S \ref{Section:3}, theoretical comparisons between the growth rates of 3-wave systems and 5-wave systems for different combinations of primary waves are provided. In \S \ref{Section:4}, numerical validations are provided for the 5-wave systems, and specific focus is also given to the fate of the primary waves near the critical latitude.  {The main results, some implications, and relevance of the settings considered in the paper are given in a brief format in \S \ref{Section:5}.}

\section{{Governing equations}} \label{Section:2}

  {The governing equations we use to study inertia-gravity waves in a rotating and stratified fluid are the 3D, incompressible, Boussinesq, Navier-Stokes equations in the $f-$plane, which in primitive variables are given by}
\begin{subequations}
 \begin{align}
    \frac{ \textnormal{D} \textbf{u}}{ \textnormal{D} t}  + f\hat{z} \times \textbf{u} &= -\frac{1}{\rho_0} \nabla p  + b \hat{z} + \nu  {\nabla^2} \textbf{u}, \label{eqn:total_mom_equation} \\
 {\frac{ \textnormal{D} b}{\textnormal{D} t} + N^2w} & {=   0},  \label{eqn:boyancy_equation}\\
    \nabla.\textbf{u} &= 0.
\label{eqn:Continuity}
 \end{align}
 \end{subequations}
Here $\textbf{u} = (u,v,w)$, where the components respectively denote the zonal, meridional, and vertical velocities. Moreover,  $f$ is the local Coriolis frequency,  $\rho_0 \approx 1000 \,\textnormal{kg}\,\textnormal{m}^{-3}$ is the reference density, $p$  is the perturbation pressure,  $b$ is the buoyancy perturbation, $\nu$ is the kinematic viscosity, and $N$ is the 
{background} buoyancy frequency.  {For simplicity, we assume the diffusion of the   {stratifying agent} to be negligible.}  
 {${ \textnormal{D} }/{ \textnormal{D} t} \equiv \partial/{\partial t} + (\textbf{u}.\nabla)$ is the material derivative.}  {$\hat{z}$ is a unit vector that points towards the positive vertical direction.} 

We intend to study wave-wave interactions using multiple-scale analysis. To this end, we combine equations \eqref{eqn:total_mom_equation}--\eqref{eqn:Continuity} into a single equation. 
After some simple manipulations, the single equation describing the evolution of vertical velocity is written in a compact form 
\begin{equation}
     \frac{\partial^2 }{\partial t^2} ( {\nabla^2} w)  + N^2(\nabla^2_h w) + f^2\frac{\partial^2 w}{\partial z^2} + \textnormal{NLT} \hspace{0.4cm}= \hspace{0.4cm} \textnormal{VT},    \label{eqn:final_w_S4}
\end{equation}
where $\nabla_h^2 \equiv \partial^2/\partial x^2 + \partial^2/\partial y^2$. $\textnormal{NLT}$ denotes all the nonlinear terms and is given by
\begin{align}
    \textnormal{NLT} &=  \nabla^2_h\frac{\partial (\textbf{u}.\nabla w)}{\partial t } + \nabla^2_h (\textbf{u}.\nabla b) - \frac{\partial^3 (\textbf{u}.\nabla u)}{\partial x \partial z \partial t } \nonumber \\
    &+     f\frac{\partial^2 (\textbf{u}.\nabla u)}{\partial y \partial z } - f\frac{\partial^2 (\textbf{u}.\nabla v)}{\partial x \partial z }- \frac{\partial^3 (\textbf{u}.\nabla v)}{\partial y \partial z \partial t }. 
    \label{eqn:NLT_s4_definition}
\end{align}
Moreover,  {$\textnormal{VT}$ denotes viscous diffusion terms and is given by}
 {\begin{equation}
    \textnormal{VT}= \nu \left[\frac{\partial}{\partial t} \left(  {\nabla^2} ( {\nabla^2} w)  \right)  + f  \frac{\partial^2 ( {\nabla^2} u)}{\partial y \partial z }   - f \frac{\partial^2 ( {\nabla^2} v)}{\partial x \partial z }\right].
    \label{eqn:VT_s4_definition}
\end{equation}}
Furthermore, we mainly focus on plane waves. Similar to the procedure used in \cite{bourget}, the vertical velocity of the $j-$th wave $(j=1,2,\ldots,5)$ is assumed to be a product of a rapidly varying phase and an amplitude that slowly varies in time. Mathematically this can be written as
\begin{equation}
      {w_j(x,y,z,t)  =   a_j(\epsilon_{t,j} t) \exp{[\textnormal{i}(k_jx + l_jy + m_jz -\omega_j t)]} + \mathrm{c.c}.,
    \label{eqn:w_definition}}
\end{equation}
where $k_j,l_j,m_j,$ and $\omega_j$ are respectively the zonal wavenumber, meridional wavenumber, vertical wavenumber, and frequency of the $j-$internal wave. {‘c.c.’ denotes the complex conjugate}.    The amplitude $a_j$ is assumed to evolve on a slow time scale $\epsilon_{t,j} $, where $\epsilon_{t,j}$ is a small parameter. Here, we assume that $\epsilon_{t,j}$ depends on $j$ because the amplitude of primary waves and the secondary waves can evolve at different time scales. {Throughout this paper, we assume steepness (defined in \S \ref{section:1}, and hereafter denoted by $\mathcal{S}_j$) is a small quantity. This is because, as explained in \S \ref{section:1}, for weakly nonlinear wave-wave interactions, $\mathcal{S}_j$ should be a small quantity \citep{koudella_staquet_2006}. Mathematically, steepness is given by
\begin{equation}
    \mathcal{S}_j \equiv \frac{ a_j m_j \sqrt{(k_j^2 + m_j^2)}}{k_j N}.
\end{equation}
The condition $\mathcal{S}_j \ll 1$ is used as a constraint for $a_j$,
\begin{equation}
    a_j  \ll  \frac{N k_j}{m_j \sqrt{(k_j^2 + m_j^2)}}.
\end{equation}
 {Substituting} \eqref{eqn:w_definition} in \eqref{eqn:final_w_S4}, at the leading order ($\mathcal{O}(\mathcal{S}_j)$) we obtain the dispersion relation in 3D, and it is given by
\begin{equation}
    \omega_j^2 = \frac{N^2(k_j^2 + l_j^2) + f^2m_j^2}{k_j^2 + l_j^2 + m_j^2}.
    \label{eqn:3d_disp_relation}
\end{equation}
All 5 waves involved in the interaction must satisfy the dispersion relation. Energy transfer between the waves due to weakly nonlinear wave-wave interactions occurs at $\mathcal{O}(\epsilon_{t}\mathcal{S},\mathcal{S}^2)$.} 

For the $j$-th wave, the amplitude evolution equation reads
\begin{align}
   \mathcal{D}_j \frac{\partial a_j}{\partial t}    &= - \textnormal{NLT}_j + \textnormal{VT}_j, 
\label{eqn:o_epsilon_2_equation_j_wave}
\end{align}
where $\mathcal{D}_j \equiv 2\textnormal{i} \omega_j (k_j^2+l_j^2+m_j^2)$ is defined for convenience. $\textnormal{NLT}_j$ and $\textnormal{VT}_j$ represent all the nonlinear and viscous terms with the phase of the $j-$th wave, respectively. The expression for $\textnormal{VT}_j$ is given by
\begin{equation}
    \textnormal{VT}_j = -\mathcal{D}_j {\frac{\nu}{2}}\left( \frac{f^2 m_j^2}{\omega_j^2} + m_j^2 + l_j^2 + k_j^2\right).
    \label{eqn:Viscous_term}
\end{equation}
$\textnormal{NLT}_j$ is obtained by substituting the fields $(u_j,v_j,w_j,b_j)$ in $\textnormal{NLT}$, and by retaining all the nonlinear terms that have the same phase as the $j-$th wave. Nonlinear terms that do not have the phase of any of the five waves are the `non-resonant terms' and are neglected. From $w_j$, we can obtain $u_j,v_j,$ and $b_j$ by using the polarisation relations, and they are given by
   {\begin{align}
    \begin{bmatrix}
u_j\\ 
v_j\\   
b_j   
\end{bmatrix} 
=    \begin{bmatrix}
U_j\\ 
V_j\\   
B_j   
\end{bmatrix}
w_j
=    \begin{bmatrix}
- {m_j(\omega_j k_j + \textnormal{i} l_j f_j)}{/[\omega_j(k_j^2 + l_j^2)]}\\ 
 - {m_j(\omega_j l_j - \textnormal{i} k_j f_j)}{/[\omega_j(k_j^2 + l_j^2)]}\\   
 -\textnormal{i} {N^2}{/\omega_j}
\end{bmatrix}
w_j.
\label{eqn:polarisation_relations}
\end{align}}
Polarisation expressions are also used to evaluate $\textnormal{NLT}_j$, the expressions for which are given in Appendix \ref{App:A}. 

\subsection{ {Wave-amplitude equations and growth rates}}

The amplitude evolution of each of the $5$ waves can be obtained from
  \eqref{eqn:o_epsilon_2_equation_j_wave}, and they are given by
\begin{equation}
    \dv{a_{1}}{t} = {\mathcal{M}}_1 a_2  a_3 - \mathcal{V}_1 a_1, \hspace{1.0cm}  \dv{a_{2}}{t} = {\mathcal{M}}_2 a_1  \bar{a}_3 - \mathcal{V}_2 a_2  {,}
    \label{eqn:triad_1_amp_evol}
\end{equation}
\begin{equation}
  \hspace{-0.4cm}  \dv{a_{5}}{t} = \mathcal{N}_5 a_4  a_3 - \mathcal{V}_5 a_5 {,}  \hspace{1.0cm}     \dv{a_{4}}{t} = \mathcal{N}_4 a_5  \bar{a}_3  - \mathcal{V}_4 a_4  {,}
    \label{eqn:wave_5_amp_evol}
\end{equation}
\begin{equation}
    \dv{a_{3}}{t} =  {\mathcal{M}}_{3} a_1  \bar{a}_2 + \mathcal{N}_{3} a_5  \bar{a}_4 - \mathcal{V}_3 a_3 {,}
    \label{eqn:wave_3_amp_evol}
\end{equation}
where $\mathcal{V}_j = {\nu}/{2}\left( {f^2 m_j^2}/{\omega_j^2} + m_j^2 + l_j^2 + k_j^2\right)$.  $\bar{a}_j$ denotes the complex conjugate of $a_j$. As shown in figure \ref{fig:5_wave_triad_schematic}(c), in this paper, we always assume that wave-1 and -5 are primary waves, and wave-2, -3, and -4 are always secondary waves.  As depicted in figure \ref{fig:5_wave_triad_schematic}(c), wave-1,-2, and -3 form a triad whose nonlinear coefficients are given by $\mathcal{M}_j$. Likewise,  wave-3, -4, and -5 also form a triad whose nonlinear coefficients are given by $\mathcal{N}_j$.  Expressions for $\mathcal{M}_j$ and $\mathcal{N}_j$ are given in  Appendix \ref{App:A}. Wave-3, therefore, becomes the common secondary wave in two different triads. 

  We now use the pump wave approximation \citep{mcewan,young} to simplify the reduced order equations \eqref{eqn:triad_1_amp_evol}--\eqref{eqn:wave_3_amp_evol}.  Under this approximation, the amplitudes of the primary waves are assumed to be constant. The assumption is only valid in the initial stages of the secondary waves' growth when the primary wave's energy does not vary significantly. In the later stages, the assumption does not give accurate results when the secondary waves' energy is comparable to the primary wave's energy. Equations \eqref{eqn:triad_1_amp_evol}--\eqref{eqn:wave_3_amp_evol} can be simplified to a set of linear differential equations which are given by
   {\begin{equation}
  \begin{bmatrix}
\dv{ \bar{a}_{2}}{t}\\ 
\dv{ \bar{a}_{4}}{t}  \\ 
\dv{ {a}_{3}}{t}   
\end{bmatrix} 
=  \begin{bmatrix}
- \mathcal{V}_2 & 0  & \bar{{\mathcal{M}}}_2 \bar{A}_1 \\ 
0 & - \mathcal{V}_4  & \bar{\mathcal{N}}_4 \bar{A}_5 \\  
{\mathcal{M}}_{3} {A}_1& {\mathcal{N}}_{3} {A}_5  & -\mathcal{V}_3 
\end{bmatrix} 
\begin{bmatrix}
\bar{a}_{2}\\ 
\bar{a}_{4}\\  
    {a}_{3}
\end{bmatrix}. 
\label{eq:scatter_matrix_scho}
\end{equation}}
 {$a_1$ and $a_5$ have been changed to $A_1$ and $A_5$, respectively, to denote the fact that they are now constants.} By assuming $ \textnormal{d} a_j/ \textnormal{d} t = \sigma a_j$, we arrive at the equation
\begin{equation}
    (\sigma + \mathcal{V}_2)(\sigma + \mathcal{V}_3)(\sigma + \mathcal{V}_4) - \bar{\mathcal{N}}_{4}{\mathcal{N}}_{3} |A_5|^2(\sigma + \mathcal{V}_2) - \bar{\mathcal{M}}_{2}{\mathcal{M}}_{3} |A_1|^2(\sigma + \mathcal{V}_4) = 0,
    \label{eqn:cubic_equation_root}
\end{equation}
where $\sigma$ is the growth rate of the system of equations given in \eqref{eq:scatter_matrix_scho}. A real, positive $\sigma$ implies that the secondary waves can extract energy from the primary wave. For $\nu = 0$ (inviscid flow), the growth rate has a simple expression given by
\begin{equation}
        \sigma = \sqrt{\bar{\mathcal{M}}_{2}{\mathcal{M}}_{3} |A_1|^2 + \bar{\mathcal{N}}_{4}{\mathcal{N}}_{3} |A_5|^2}.
        \label{eqn:growth_rate_inviscid}
\end{equation}
{The derivation of equation \eqref{eqn:growth_rate_inviscid} is given in Appendix \ref{App:B}.}   {We arrive at the standard growth rate expression for triads (3-wave systems) by setting either $A_1=0$ or $A_5=0$.}
   Moreover, we can also obtain the condition
\begin{equation}
        \sqrt{\bar{\mathcal{M}}_{2}{\mathcal{M}}_{3} |A_1|^2 + \bar{\mathcal{N}}_{4}{\mathcal{N}}_{3} |A_5|^2} \leq \sqrt{2} \widehat{\sigma}_{1} \hspace{0.5cm} \textnormal{or} \hspace{0.5cm}         \sqrt{\bar{\mathcal{M}}_{2}{\mathcal{M}}_{3} |A_1|^2 + \bar{\mathcal{N}}_{4}{\mathcal{N}}_{3} |A_5|^2} \leq \sqrt{2} \widehat{\sigma}_{5}
        \label{eqn:growth_rate_inequality}
\end{equation}
where $\widehat{\sigma}_1(\widehat{\sigma}_5)$ is the maximum growth rate of all 3-wave systems of primary wave-1(5). If both the primary waves have the same amplitude $(A_1=A_5)$, frequency, and wavevector norm, then $\widehat{\sigma}_{1}=\widehat{\sigma}_{5}$. In such cases,  \eqref{eqn:growth_rate_inequality} implies that a 5-wave system's growth rate could, in principle, be higher ({the} maximum being $\sqrt{2}$ times) than the maximum growth rate of all 3-wave systems.  In this paper, we consistently assume $A_1=A_5$.

\subsection{{5-wave system identification}} \label{sec:2.2}
For a resonant 5-wave system, all three secondary waves should satisfy the dispersion relation. This leads to 3 constraints, where $(\omega_j,k_j,l_j,m_j)$ for $j=2,3,4$ have to satisfy equation \eqref{eqn:3d_disp_relation}.
The triad conditions also add additional constraints, and they are given by
\begin{subequations}
 \begin{align}
    \omega_2 &= \omega_1 - \omega_3, \hspace{0.3cm} \textbf{k}_2 =  \textbf{k}_1 - \textbf{k}_3, \label{eqn:triad_cond_wv_2} \\
    \omega_4 &= \omega_5 - \omega_3, \hspace{0.3cm} \textbf{k}_4 =  \textbf{k}_5 - \textbf{k}_3, \label{eqn:triad_cond_wv_4} 
\end{align}   
\end{subequations}
where $\textbf{k}_j = (k_j,l_j,m_j)$ is the wavevector of the $j-$wave. Using equations \eqref{eqn:triad_cond_wv_2} and \eqref{eqn:triad_cond_wv_4} respectively in the dispersion relation of wave-2 and wave-4 leads to
\begin{subequations}
 \begin{align}
(\omega_1-\omega_3)^2 &= \frac{N^2((k_1-k_3)^2 + (l_1-l_3)^2) + f^2(m_1-m_3)^2}{(k_1-k_3)^2 + (l_1-l_3)^2 + (m_1-m_3)^2},   \label{eqn:Final_Id_2} \\
(\omega_5-\omega_3)^2 &= \frac{N^2((k_5-k_3)^2 + (l_5-l_3)^2) + f^2(m_5-m_3)^2}{(k_5-k_3)^2 + (l_5-l_3)^2 + (m_5-m_3)^2}.   \label{eqn:Final_Id_3} 
\end{align}   
\end{subequations}
Solutions for \eqref{eqn:Final_Id_2}-\eqref{eqn:Final_Id_3}, along with the solution for equation \ref{eqn:3d_disp_relation} for $j=3$, would provide resonant 5-wave systems, and they are found by varying $(\omega_3,k_3,l_3,m_3)$.

Hereafter we always assume $|\textbf{k}_1| =  |\textbf{k}_5|$, however, $\textbf{k}_1 \neq  \textbf{k}_5$, and $\omega_1 = \omega_5=0.1N$. Such small frequency values appear in many other studies,  for example,  \cite{sonya_nik_2011,Mathur_14}.   {Moreover, we consistently use $f<N$.  {There is an abundance of internal waves in the ocean satisfying the relation $f<\omega_j<N$} (see figure 4 of \cite{zhao}).  {In this study, we only focus on this regime.}
The equations we derived should also be applicable for the parameter regime $f > N$ as long as $f > \omega_j > N$ is satisfied. Triad interactions for this parameter regime have been studied recently \cite{inertial_waves_triad}.}

\section{{Results from the reduced--order model}} \label{Section:3}

\subsection{ Primary waves in the same vertical plane}

\subsubsection{  { {Case} $\mathbf{k}_1 =  (k_1,0,m_1)$ and $\mathbf{k}_5 =  (k_1,0,-m_1)$ } } \label{sec:m_minus_m}

We first consider the scenario where the two primary waves have the same horizontal wavevector ($k,l$) but propagate in vertically opposite directions, see figure \ref{fig:5_wave_triad_schematic}(a).  For simplicity,  the meridional wavenumbers of the primary waves ($l_1$ and $l_5$) are assumed to be 0.
Internal waves propagating in vertically opposite directions are ubiquitous in the oceans. For example, internal wave beams getting reflected from the bottom surface of the ocean, from the air-water interface, or even from the pycnocline will result in scenarios where primary waves propagating in vertically opposite directions meet. 

For the given set of primary waves, a resonant 5-wave system is possible only when $\omega_3\approxeq \omega_1-f$. No other resonant 5-wave systems were found for $0<f/\omega_1<0.5$. Hence, the 5-wave system always consists of (a) two primary waves, each with frequency $\omega_1$ (as per our assumption), (b) a common secondary wave with frequency $\omega_1-f$,  and (c) two near-inertial (frequency $\approx f$) secondary waves, which propagate in vertically opposite directions. 

Next, we study the growth rates of the 5-wave system. First, we decide on the viscosity values in a non-dimensionalised form. In this regard, we use $|A_1|/k_1\nu = 10^{4}$ and $|A_1|/k_1\nu = 10^{7}$ throughout the paper.  At $|A_1|/k_1\nu = 10^{7}$, viscous effects are usually negligible, hence $|A_1|/k_1\nu = 10^{4}$ is also considered to see what 5-wave systems are affected by viscosity. We note in passing that $|A_1|/k_1\nu \sim \mathcal{O}(10^{6})$ was used by \cite{bourget} to study triads with realistic oceanic parameters.  

\begin{figure}
\centering{\includegraphics[width=1.0\textwidth]{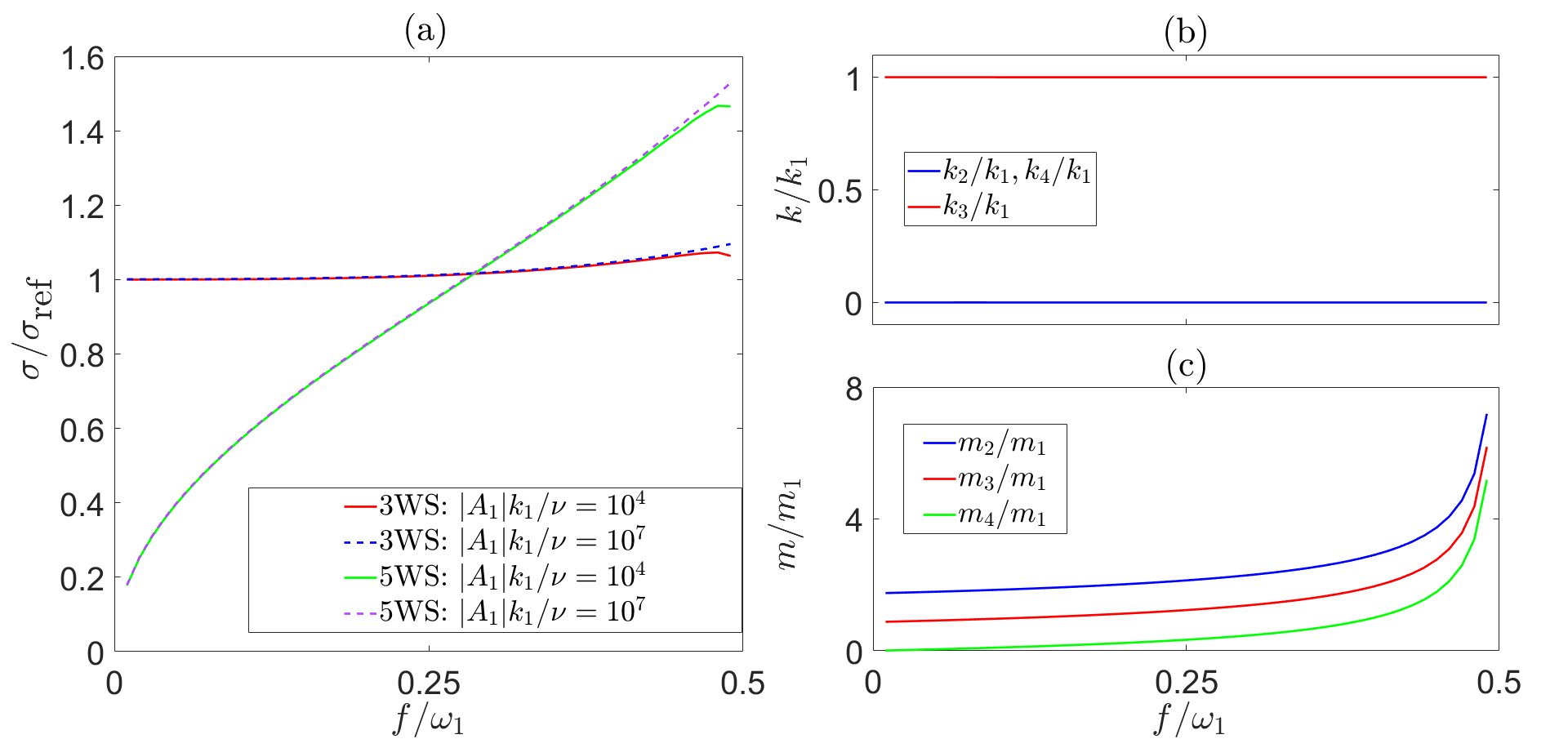}}
 \caption{ (a) Variation of 5-wave system's (denoted by 5WS) growth rate and 3-wave systems' (denoted by 3WS) maximum growth rate with $f$ for $\mathbf{k}_1 =  (k_1,0,m_1)$ and $\mathbf{k}_5 =  (k_1,0,-m_1)$ for two different viscosity values. $\sigma_{\textnormal{ref}}$ is the maximum growth rate of a 3-wave system at $f/\omega_1=0.01$ and $|A_1|/k_1\nu = 10^{4}$.
 The horizontal and vertical wavenumbers of the secondary waves in the 5-wave system are respectively shown in (b) and (c).  {In (b),  $k_2=k_4=0$.}   }
\label{fig:k_m_k_minus_m_max_growth_rate_1p_vs_2p}
\end{figure}

Figure \ref{fig:k_m_k_minus_m_max_growth_rate_1p_vs_2p}(a) shows how the maximum growth rate of the 5-wave system and 3-wave systems vary with $f/\omega_1$ for different $\nu$. The growth rates are evaluated by fixing $k_1$ and $A_1$ as $f$ is varied. Figures \ref{fig:k_m_k_minus_m_max_growth_rate_1p_vs_2p}(b)--\ref{fig:k_m_k_minus_m_max_growth_rate_1p_vs_2p}(c)  respectively show the horizontal  and the vertical wavenumbers of the secondary waves involved in the 5-wave interaction. Note that the common secondary wave's horizontal wavevector  $(k_1,0)$ is always the same as that of the primary waves.  This is expected since the other two secondary waves are near-inertial waves. 

Moreover, the common secondary wave can have a positive or a negative vertical wavenumber. This arises from the symmetry of the primary waves' vertical wavenumber: there is no reason that $m$ is more special than $-m$ or vice versa. As expected, the growth rates only depend on the magnitude of the common secondary wave's vertical wavenumber. For low $f$ values, figure \ref{fig:k_m_k_minus_m_max_growth_rate_1p_vs_2p} shows the 3-wave system has a higher growth rate, hence it is the dominant instability. This is because the two 3-wave systems that combine to form the 5-wave system always contain near-inertial secondary waves. Moreover, the growth rate of 3-wave systems containing near-inertial waves is much smaller than the maximum possible growth rate \cite[figure 8]{richet_2018} at low latitudes. As a result, the resonant 5-wave system, which is a combination of two 3-wave systems (as shown in equation \ref{eqn:growth_rate_inviscid}), is of little significance at low latitudes.

 As $f$ increases, the 5-wave system's growth rate becomes higher than the maximum growth rate of all 3-wave systems. The transition occurs near $f/\omega_1 \approx 0.3$, see figure \ref{fig:k_m_k_minus_m_max_growth_rate_1p_vs_2p}(a).  For high values of $f/\omega_1$, 5-wave systems may be  {more dominant} in locations where an internal wave beam gets reflected from a flat bottom surface or from a nearly flat air-water surface. 
However, for inclined reflecting surfaces, the results presented here (which are based on the assumption that the two primary waves have the same wavevector norm) may not be valid since inclination results in a significant change in wavevector norm \citep{phillips}. 

Finally, it can also be observed that the primary wave combination considered in this subsection produces a field that resembles an internal wave mode in a vertically bounded domain. Hence, the predictions made in this section should also hold for modes in a bounded domain. However, in a vertically bounded domain, only a discrete set of vertical wavenumbers is allowed for a particular frequency. As a result, for a resonant 5-wave system to exist, the vertical wavenumbers of the secondary waves should be a part of the discrete vertical wavenumber spectrum. 

\subsubsection{   { {Case} $\mathbf{k}_1 =  (k_1,0,m_1)$ and $\mathbf{k}_5 =  (-k_1,0,m_1)$ } } 
\label{sec:horizontally_opposite_P_waves} 

\begin{figure}
 \centering{\includegraphics[width=1.0\textwidth]{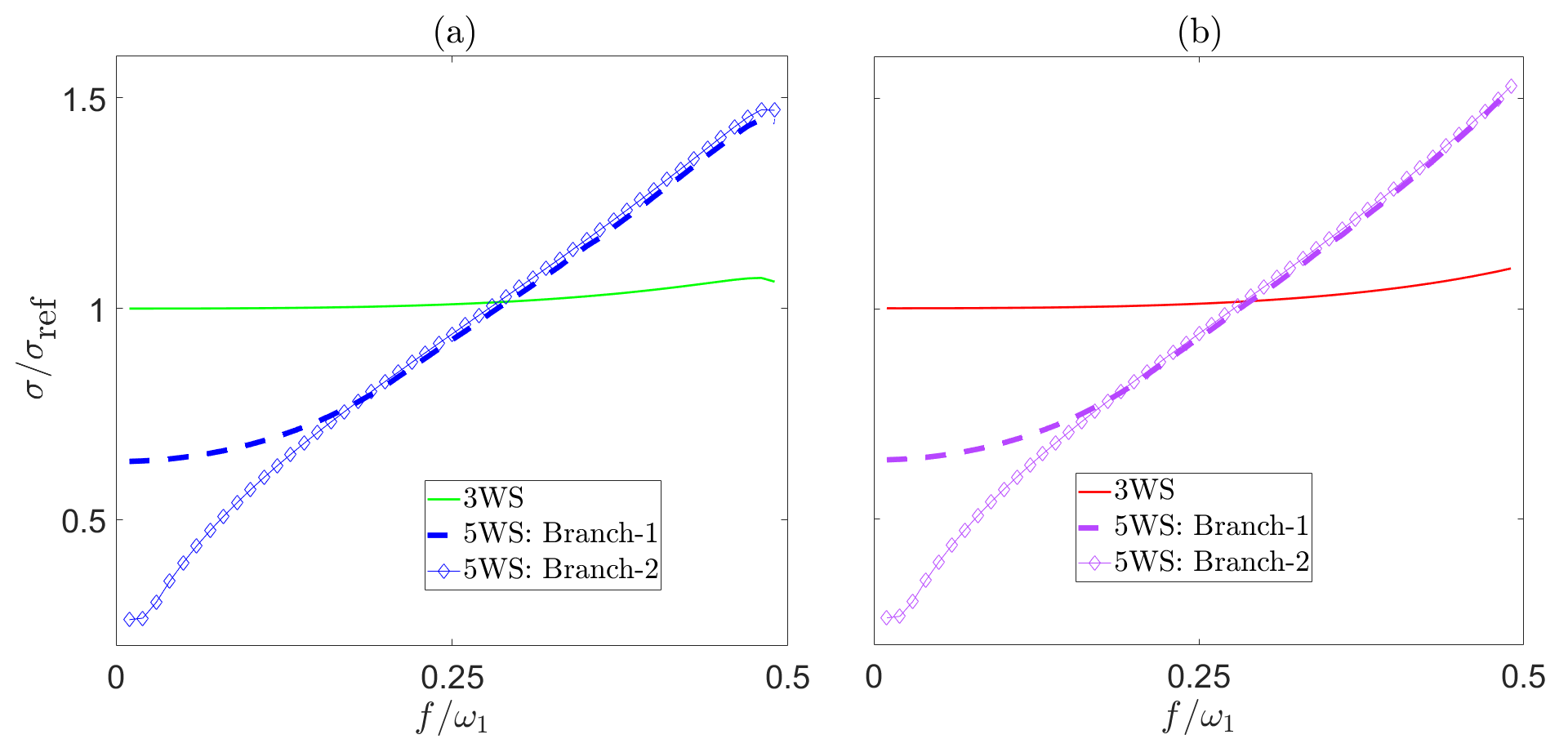}}
 \caption{ Comparison of maximum growth rates of 5-wave systems and 3-wave systems for $\mathbf{k}_1 =  (k_1,0,m_1)$ and $\mathbf{k}_5 =  (-k_1,0,m_1)$.  (a) $|A_1|/k_1\nu = 10^{4}$, and (b) $|A_1|/k_1\nu = 10^{7}$.  }
  \label{fig:k_m_minus_k_m_max_growth_rate_1p_vs_2p}
\end{figure}

Here we focus on the scenario where the two primary waves have the same vertical wavenumber but propagate in horizontally opposite directions, as given in figure \ref{fig:5_wave_triad_schematic}(b). Moreover, $l_1=l_5=0$ is again assumed. For this particular combination of primary wavevectors, resonant 5-wave systems are possible for $\omega_3 \in (f, 0.53\omega_1)$. For $f/\omega_1 = 0.01$, resonant 5-wave systems exist up to $\omega_3 \approx 0.53\omega_1$. As $f$ increases, the maximum possible value of $\omega_3$ slowly reduces to $0.5\omega_1$. 
We define two branches: 
5-wave systems where the common secondary wave has a positive (negative) vertical wave number are defined as Branch-1(2). 
Figures \ref{fig:k_m_minus_k_m_max_growth_rate_1p_vs_2p}(a)--\ref{fig:k_m_minus_k_m_max_growth_rate_1p_vs_2p}(b) show how the maximum growth rate for each of these two branches varies with $f$ for two different viscosity values. The maximum growth rate of 3-wave systems is once again plotted to provide a clear comparison between 5-wave and 3-wave systems. For lower $f$ values, 5-wave systems have a lesser maximum growth rate than the maximum growth rate of 3-wave systems ($\sigma/\sigma_{\textnormal{ref}}<1$). However, the 5-wave instability is more dominant than the 3-wave instability for higher $f$ values. The transition once again occurs near $f \approx 0.3\omega_1$.  All these observations are similar to those in figure \ref{fig:k_m_k_minus_m_max_growth_rate_1p_vs_2p}(a). For high $f$ values, the maximum growth rate of both branches is almost the same.
Viscosity has a non-negligible effect only when $f\approx\omega_1/2$, where the secondary waves have a high vertical wavenumber.

\begin{figure}
 \centering{\includegraphics[width=1.0\textwidth]{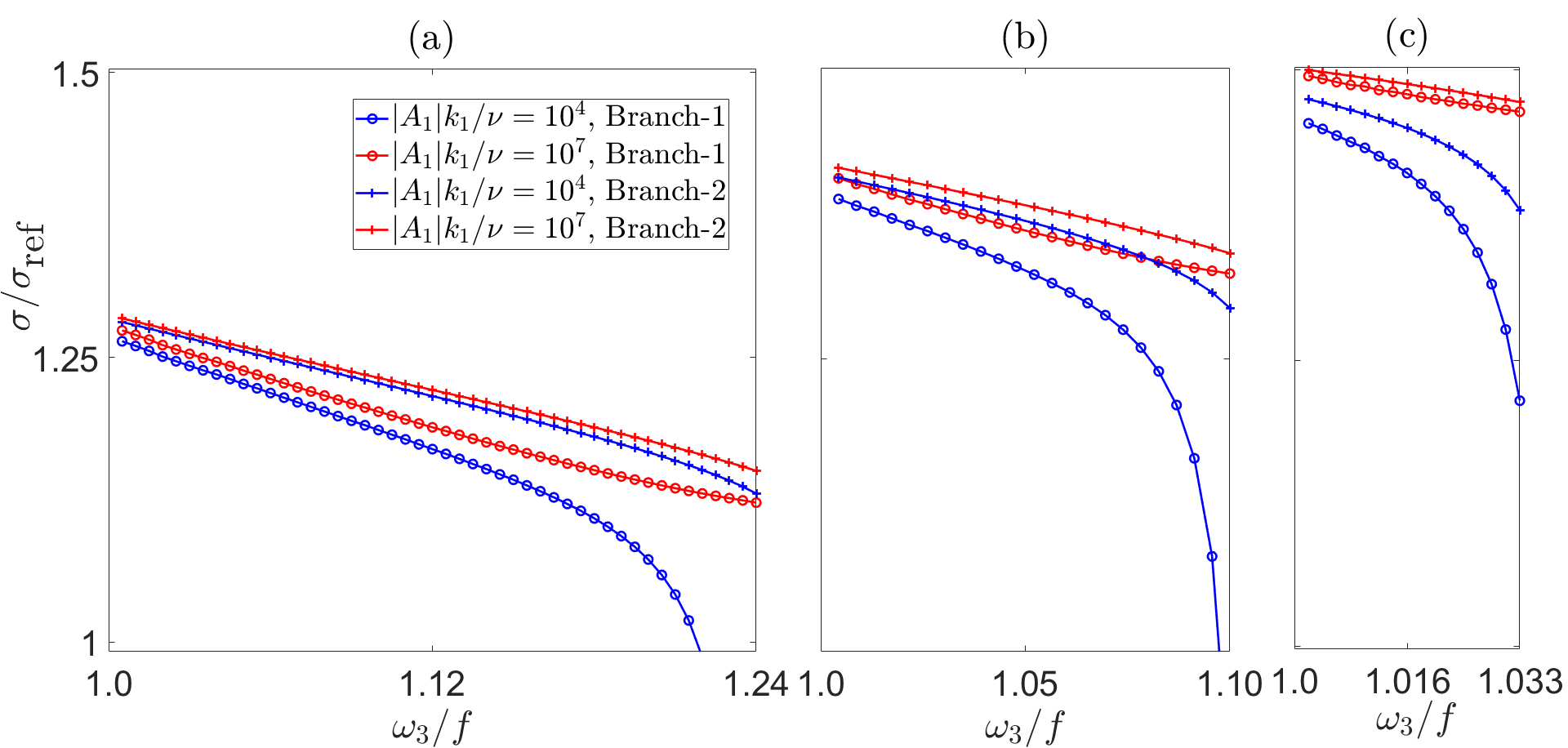}}
 \caption{ Growth rate variation with $\omega_3/f$ for Branch-1 and 2 for (a) $f/\omega_1 = 0.40$, (b) $f/\omega_1 = 0.45$, and (c) $f/\omega_1 = 0.48$. For all the plots, the maximum value of abscissa ($\omega_3/f$) corresponds to $\omega_3/\omega_1 \approx 0.496$. 
 }
  \label{fig:B12_k_m_minus_k_m}
\end{figure}

Figures \ref{fig:B12_k_m_minus_k_m}(a)--\ref{fig:B12_k_m_minus_k_m}(c) show how the growth rate of both the branches vary with $\omega_3/f$ for three different $f/\omega_1$ values.   {The growth rates always decrease as $\omega_3$ is increased, indicating the maximum growth rate is at $\omega_3=f$.} For $f/\omega_1 > 0.3$, the common secondary wave is always a near-inertial wave in the most unstable 5-wave system. Interestingly, as $\omega_3$ is increased from $f$, the meridional wavenumber of the common secondary wave increases, hence making the instability 3D. Moreover, for the three $f$ values analysed in figure \ref{fig:B12_k_m_minus_k_m}, the zonal wavenumber of the common secondary wave $(k_3)$ is nearly zero for all the 5-wave systems. The maximum growth rate occurs at $\omega_3 \approx f$ where $(k_3,l_3) \xrightarrow{} 0$. As a result, the system's most unstable mode can be studied/simulated by considering a 2D system. 
The effects of viscosity are more apparent for $|A_1|/k_1\nu = 10^{4}$ as expected, and Branch-1 is affected by viscous effects more than Branch-2.

For high $f$ values, near-inertial waves have been observed to be the secondary waves of a primary internal wave with semidiurnal frequency, see \cite{legg_2017_dissipation_generation,mean_current_richet_triad,richet_2018,GM_spectrum_tidal_forcing}. In the case of topographic
generation of internal waves, internal wave beams intersecting each other is quite common. The locations where internal wave beams intersect can serve as spots where a single near-inertial wave can extract energy from two different internal wave beams. 

\subsection{Oblique primary waves}

\begin{figure}
 \centering{\includegraphics[width=1.0\textwidth]{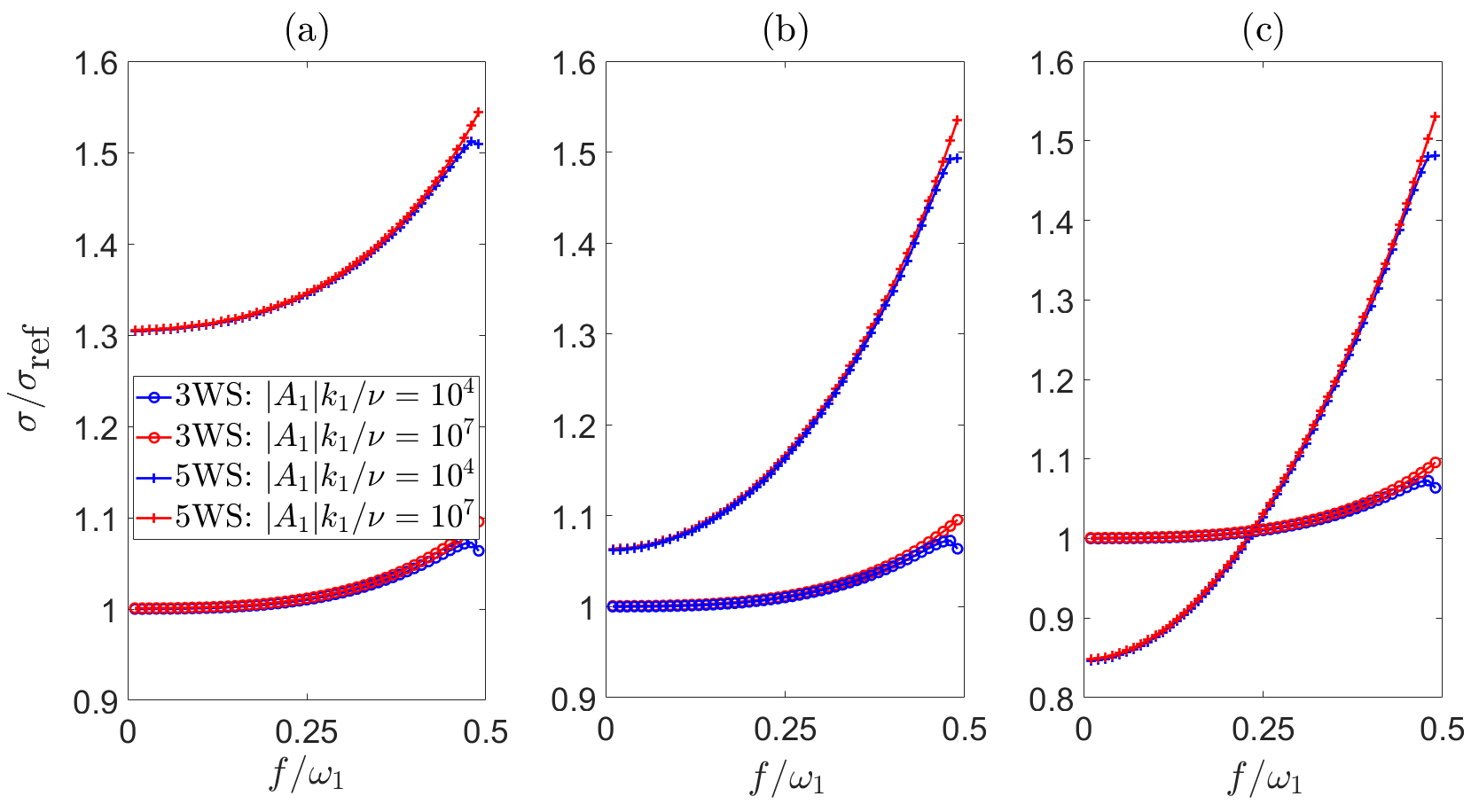}}
 \caption{ Variation of maximum growth rate with $f$ for 5-wave systems and 3-wave systems for an oblique set of primary waves.  (a) $\theta = \pi/4$, (b) $\theta = \pi/2$, and (c) $\theta = 3\pi/4$. }
  \label{fig:max_growth_rate_oblique_1p_vs_2p}
\end{figure}

In the oceans, primary waves that are not on the same vertical plane can also propagate amidst each other. Here we study the maximum growth rate for 5-wave systems where the primary waves have a non-zero meridional wavenumber. The primary wavevectors are given by 
\begin{equation}
    \mathbf{k}_1 =  (k_1\sin{(\theta/2)},k_1\cos{(\theta/2)},m_1), \hspace{1cm}  \mathbf{k}_5 =  (-k_1\sin{(\theta/2)},k_1\cos{(\theta/2)},m_1),
    \label{eqn:theta_definition}
\end{equation}
  {where the parameter $\theta$ is the angle
between the two primary wavevectors in the $(k, l)$ plane}. Note that $\theta = \pi$ leads to the wavevector combination $\mathbf{k}_1 =  (k_1,0,m_1)$ and $\mathbf{k}_5 =  (-k_1,0,m_1)$ considered in \S \ref{sec:horizontally_opposite_P_waves}.  Following \eqref{eqn:theta_definition}, the condition $|\mathbf{k}_1| = |\mathbf{k}_5|$ will be automatically satisfied.
 {We analyse how the growth rate of 5-wave systems varies with $\theta$.}
Figures \ref{fig:max_growth_rate_oblique_1p_vs_2p}(a)--\ref{fig:max_growth_rate_oblique_1p_vs_2p}(c) show the variation of the maximum growth rate of 3-wave systems and 5-wave systems with $f$ respectively for three different $\theta$ values:  $\pi/4$, $\pi/2$, and $3\pi/4$. 
Increasing $\theta$ results in 5-wave systems being less effective than 3-wave systems in the lower latitudes. For $\theta=\pi/4$, the 5-wave system is the dominant instability regardless of the latitude. A similar result is observed for $\theta=\pi/2$, however, the difference between the 5-wave and 3-wave systems is clearly reduced compared to $\theta=\pi/4$. For $\theta=3\pi/4$, the 5-wave system is the dominant instability only for $f/\omega_1 \gtrapprox 0.25$.  {Considering the results from \S \ref{sec:horizontally_opposite_P_waves}, for any $\theta$,  {5-wave instability is expected to be more dominant than the 3-wave instability} when $f/\omega_1 \gtrapprox 0.3$.}

 For $\theta = \pi$, and $f/\omega_1 \gtrapprox 0.3$, the maximum growth rate for 5-wave systems occurs when $\omega_3 = f$.
 However, for $\theta = \pi/4$ and $\pi/2$, in the most unstable 5-wave system, wave-3 is not a near-inertial wave even for $f/\omega_1 = 0.45$. Hence, as $\theta$ is reduced, the most unstable 5-wave system may not contain near-inertial waves.  {The predictions for the 5-wave system will fail as $\theta\xrightarrow{}0$ since both primary waves will have the same wavevector.}

\section{Numerical simulations} \label{Section:4}

Here we present results from numerical simulations conducted to validate the predictions from reduced-order analysis presented in \S \ref{Section:3}, with the primary focus being on \S \ref{sec:m_minus_m} and \S \ref{sec:horizontally_opposite_P_waves}. Equations \eqref{eqn:total_mom_equation}--\eqref{eqn:Continuity} are solved with the open source pseudo-spectral code Dedalus \citep{Dedalus}. In all the numerical simulations, periodic boundary conditions are used in the horizontal direction. In the vertical direction, we use periodic boundary conditions for the simulations where the primary wave is a plane wave. Moreover, we use impenetrable ($w=0$), free-slip ($\textnormal{d}u/dz,\textnormal{d}v/dz=0$) boundary condition in the vertical direction for the simulations where the primary wave is a mode-1 or a mode-2 wave. We decide the amplitude of the primary wave by fixing the maximum value of zonal velocity ($u$). Using polarisation relations, $(u_j,v_j,w_j,b_j)$ for the primary waves can be easily found out, and the other fields are also initialised at $t=0$. For numerical validations, we only consider 2D situations, i.e. $\partial/\partial y = 0$, implying $l_j=0$. The details of the simulations are as follows: we fix the primary waves' horizontal wavenumber at $k_1 = 1/H$, where   $H=500$ m. We consistently use $N = 10^{-3}$ $\textnormal{s}^{-1}$ and $\omega_1/N = 0.1$. However, we vary $f/\omega_1$, and hence the vertical wavenumber of the primary waves ($m_1$) also varies since $m_1$ is a function of $f/\omega_1$. The amplitude of the primary waves is chosen such that the maximum zonal velocity ($u$) is always $0.001$  $\textnormal{ms}^{-1}$. Computational time is variable and depends on the simulation in question. For all simulations, $4$-th order Runge-Kutta method is used as the time-stepping scheme with a time step size of $(2\pi/\omega_1)/800$ (i.e. $800$ steps for one time period of the primary wave).
 {All the fields are expressed using Fourier modes in the horizontal direction, and either 64 or 128 modes are used per one horizontal wavelength of the primary wave.  {Moreover, the vertical direction is resolved using Chebyshev polynomials (for the vertically bounded domain) or Fourier modes (for the double periodic domain), and the resolution is varied from a minimum of 96 to a maximum of 512 grid points per one vertical wavelength of the primary wave.}} All simulations are initialised with a small amplitude noise, the spectrum of which is given by
\begin{equation}
    \mathcal{R}_{\textrm{noise}}(x,z) = \int_{0}^{k_{\textrm{noise}}}\int_{ m_{\textrm{lowest}}}^{m_{\textrm{noise}}} A_{\textrm{noise}} \sin( kx + mz + \phi_{\textrm{noise}}(k,m))   { \textnormal{d}m\textnormal{d}k},
    \label{eqn:noise_definition}
\end{equation}
where $\phi_{\textrm{noise}}(k,m)\in [0,2\pi]$ is the random phase part, which   is generated using the `rand' function in Matlab for each $(k,m)$. 
Unless otherwise specified, $k_{\textrm{noise}} = 48k_1$ and $m_{\textrm{noise}} = 48m_1$. Moreover, $m_{\textrm{lowest}} = 2\pi/Lz$, where $Lz$ is the length of the domain in the $z$-direction. Equation \eqref{eqn:noise_definition} is added to the $b$ or $v$ field. The noise amplitude $A_{\textrm{noise}}$ is at least $10^{-3}$ times smaller than the primary waves' corresponding amplitude.  Unless otherwise mentioned,   {$\nu = 10^{-6}$  $\textnormal{m}^2\textnormal{s}^{-1}$} is taken.

\subsection{{  {Case $\mathbf{k}_1 =  (k_1,0,m_1)$ and $\mathbf{k}_5 =  (k_1,0,-m_1)$ }  }} \label{sec_4_1}
\begin{figure}
\centering{\includegraphics[width=1.0\textwidth]{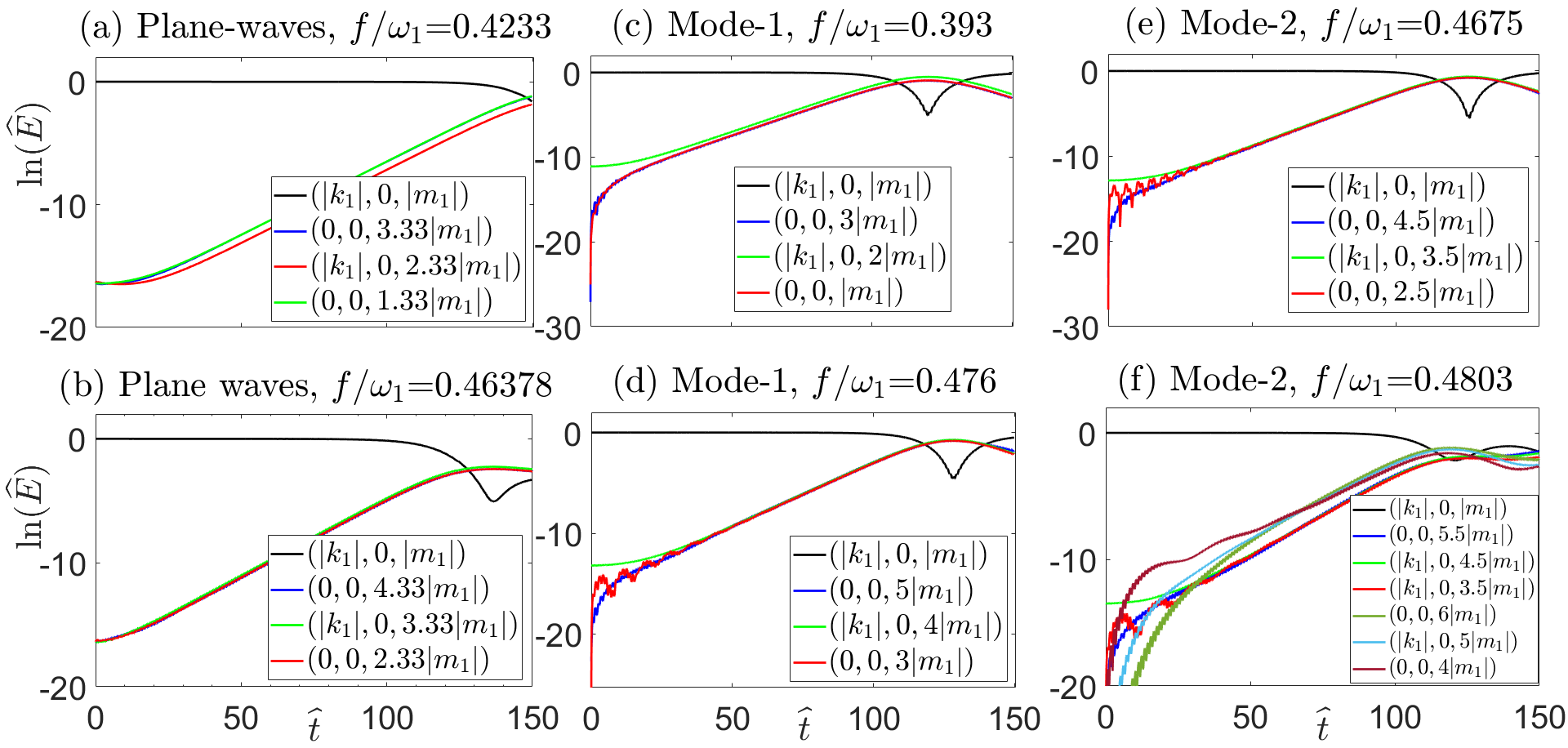}}
  \caption{ 5-wave interactions for plane waves, mode-1, and mode-2 different $f$ values (i.e. latitudes). (a) and (b) show plane waves, (c) and (d) show Mode-1 waves, and (e) and (f) show Mode-2 waves. $\widehat{t} \equiv t\omega_1/2\pi $. }
  \label{fig:modes_log_energy}
\end{figure}

\begin{figure}
 \centering{\includegraphics[width=1.0\textwidth]{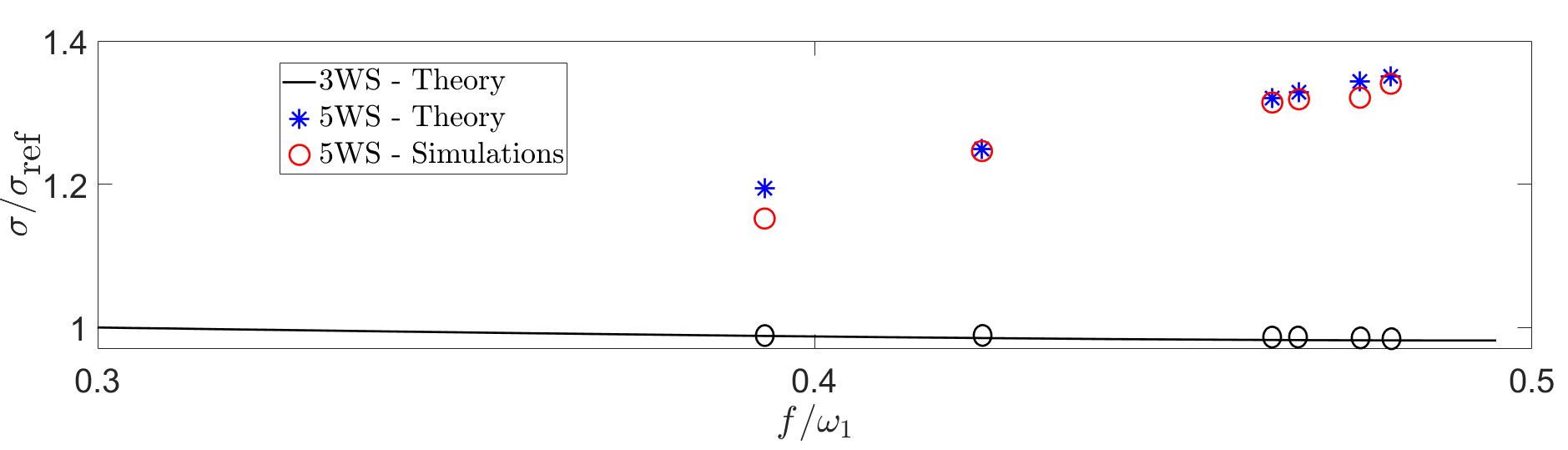}}
 \caption{ Comparison between theoretical growth rates and growth rates obtained from the simulations for $\mathbf{k}_1 =  (k_1,0,m_1)$ and $\mathbf{k}_5 =  (k_1,0,-m_1)$. Red (blue) markers indicate results from the simulations (theory). The black curve plots the variation of the maximum growth rate of 3-wave systems with $f$. }
  \label{fig:Simulation_vs_theory_modes}
\end{figure}

We first focus on the primary wavevector combination $\mathbf{k}_1 =  (k_1,0,m_1)$ and $\mathbf{k}_5 =  (k_1,0,-m_1)$.  As mentioned previously, the combination of wavevectors $\mathbf{k}_1 =  (k_1,0,m_1)$ and $\mathbf{k}_5 =  (k_1,0,-m_1)$ leads to fields that are very similar to an internal wave mode in a vertically bounded domain. As a result, we also simulate low modes (modes-1 and 2) in a vertically bounded domain to observe whether there is an emergence of the `5-wave  instability'. The decay of the primary waves is simulated at specific latitudes where the secondary waves' vertical wavenumbers in the resonant 5-wave system are multiples of $m_1/3$ or $m_1/2$. This choice helps in reducing the computational resources required for the simulations. To estimate the energy in different wavevectors, we simply use 
 the Fast Fourier Transform (FFT) for both $x$ and $z$ directions in simulations where the primary waves are plane waves. In a vertically bounded domain, FFT is used only in the $x$ direction, while for the $z-$direction,  the orthogonal nature of the modes is exploited. As a measure of the energy contained in a wavevector, a non-dimensionalised energy $\widehat{E}$ is introduced and is defined as
\begin{equation}
    \widehat{E}(k,0,m,t) = \frac{|\hat{u}(k,0,m,t)|^2 + |\hat{w}(k,0,m,t)|^2 + |\hat{v}(k,0,m,t)|^2 + |\hat{b}(k,0,m,t)|^2/N^2}{ E_{\textnormal{ref}} }  {,} 
\end{equation}
 {where the hat variables ($\hat{u},\hat{w},\hat{v},\hat{b}$) denote the Fourier amplitudes of (${u},{w},{v},{b}$), respectively.} $E_{\textnormal{ref}}$ serves as the measure of primary waves' energy at $t=0$ and is defined as 
\begin{equation}
   E_{\textnormal{ref}} =  \left(|\hat{u}(k_1,0,m_1)|^2 + |\hat{w}(k_1,0,m_1)|^2 + |\hat{v}(k_1,0,m_1)|^2 + |\hat{b}(k_1,0,m_1)|^2/N^2\right)\bigg|_{t=0}
   \label{eqn:E_ref_definition} {.}
\end{equation} 
We simulate 6 cases: 2 cases for primary waves in an unbounded domain (plane waves), and 2 cases each for mode-1 and mode-2 waves in a vertically bounded domain. For mode-1, $m_{\textrm{noise}}=96 m_1$ is chosen. For every simulation, a different $f$ value is used, and hence the resonant 5-wave system is different in each case. Figure \ref{fig:modes_log_energy} shows the exponential growth of secondary waves at  6 different latitudes due to 5-wave interactions. Columns 1, 2, and 3 of figure \ref{fig:modes_log_energy} show the results of the simulations that focus on plane waves, mode-1 waves, and mode-2 waves.
Figures \ref{fig:modes_log_energy}(a)--\ref{fig:modes_log_energy}(e) plot four different wavevectors. The wavevector $(|k_1|,0,|m_1|)$ contains the energy of both primary waves, while the other three wavevectors indicate the secondary waves.   {Even though there are only three different secondary wavevectors, two different 5 wave-systems (that are qualitatively very similar) exist in these cases. The norm of the wavevectors is the same. However, the direction of propagation of the secondary waves is different. For example,   {in figure \ref{fig:modes_log_energy}(a)}, the two different 5-wave systems are
\begin{itemize}
    \item System-1 (secondary waves): $(0,0,3.33m_1)$, $(k_1,0,-2.22m_1)$, $(0,0,1.11m_1)$ {,}
    \item System-2 (secondary waves): $(0,0,-3.33m_1)$, $(k_1,0,2.22m_1)$, $(0,0,-1.11m_1)$ {.} 
\end{itemize}
The primary waves are the same for both systems. The sign (and only the sign) of the vertical wavenumber changes between the two systems. Theoretically, the growth rates are the same for both systems.} All three secondary waves grow exponentially, which provides clear evidence that this is a 5-wave system. In \ref{fig:modes_log_energy}(f), two different 5-wave systems emerge, and both systems are plotted.  We can observe that as the $f$ value increases, the secondary waves' vertical wavenumber also increases in the simulations (see the legend), which is in line with the theoretical predictions given in figure \ref{fig:k_m_k_minus_m_max_growth_rate_1p_vs_2p}(c). In all six simulations, near-inertial waves are present ($k=0$).  The growth rate of the secondary waves is calculated by estimating $\textnormal{d}\ln{(\widehat{E})}/\textnormal{d}t$. The comparison of growth rates from simulations and theory is presented in figure \ref{fig:Simulation_vs_theory_modes}, which shows a reasonably good agreement. For all the cases, the average of the three secondary waves' growth rate in a particular 5-wave system is taken. 
Moreover, figure \ref{fig:Simulation_vs_theory_modes} reveals that the growth rates are well above the maximum growth rate of all 3-wave systems.  It can be noticed that growth rates for 3-wave systems increase in section \ref{Section:3}, but decrease in figure \ref{fig:Simulation_vs_theory_modes} as $f$ increases. In section \ref{Section:3}, the growth rates are always evaluated by fixing $k_1$ and $A_1$ as $f$ is varied. This means the vertical velocity of the primary wave is fixed as $f$ is varied, while the   {Coriolis} frequency is varied. However, in section \ref{Section:4}, we fix the amplitude of the primary wave in a different way: the horizontal velocity of the primary wave is assumed to be a constant even as $f$ is varied, and this means the quantity $A_1m_1$ is a constant. This is the reason the growth rate of 3-wave systems varies differently with $f$ in the two sections.
5-wave interactions can happen for standing modes only at specific latitudes because the vertical wavenumbers are discrete, not continuous. However, for plane waves, there is no such constraint. As per the predictions in \S \ref{Section:3},  5-wave interactions should be faster than 3-wave interactions provided $f/\omega_1 \gtrapprox 0.3$. 

In triads, the direction of energy transfer can be reversed. After some time, the secondary waves completely drain the energy of the primary wave, and from this point, the secondary waves start giving the energy back to the primary wave. This can be seen by solving the theoretical reduced order model equations. We observe this reversal for cases where the secondary waves have low wavenumbers in simulations (see figure \ref{fig:modes_log_energy}(c)--(f)). In realistic oceanic scenarios, the secondary waves themselves can behave as primary waves once they have significant energy and give their energy to smaller length scale waves, thus causing a continuous energy cascade. We believe this would occur when the secondary waves have much higher wavenumbers than the primary wave.

It was observed that as $f/\omega_1 \xrightarrow[]{} 0.5$, multiple secondary wave combinations grow and extract a considerable amount of energy from the primary waves. This can even be seen in figure  \ref{fig:modes_log_energy}(f), where two different 5-wave systems emerge and extract a significant amount of energy. As $f/\omega_1 \xrightarrow[]{} 0.5$, multiple 5-wave systems can become coupled and grow at a rate that is faster than any single 5-wave system (discussed in detail in \S \ref{sec_4_2}). Hence, the growth rates predicted from a single 5-wave interaction will not be accurate when $f \approx \omega_1/2$. As $f \xrightarrow[]{} \omega_1/2$, the growth rate for a mode-1 wave with zonal velocity amplitude $0.002\textnormal{ms}^{-1}$ will approach $2\sigma_{\textnormal{cl}}$ instead of $\sqrt{2}\sigma_{\textnormal{cl}}$, where $\sigma_{\textnormal{cl}}$ is the maximum growth rate for a plane wave with zonal velocity amplitude $0.001\textnormal{ms}^{-1}$ at the critical latitude \citep{young}.
  {{Although in} \cite{young}, the mode-1 wave was considered in the presence of a non-constant $N$,  their prediction is still expected to hold for our case (constant $N$).} Our numerical simulations also show the growth rates of the secondary waves being well above $\sqrt{2}\sigma_{\textnormal{cl}}$ for $f\approx\omega_1/2$.   We conducted simulations of a mode-1 wave undergoing wave-wave interaction instability near the critical latitude ($f/\omega_1 = 0.497$ and $0.498$ were chosen). In both simulations, secondary waves with growth rates $\gtrapprox 1.84\sigma_{\textnormal{cl}}$ were observed. 

\begin{figure}
 \centering{\includegraphics[width=1.0\textwidth]{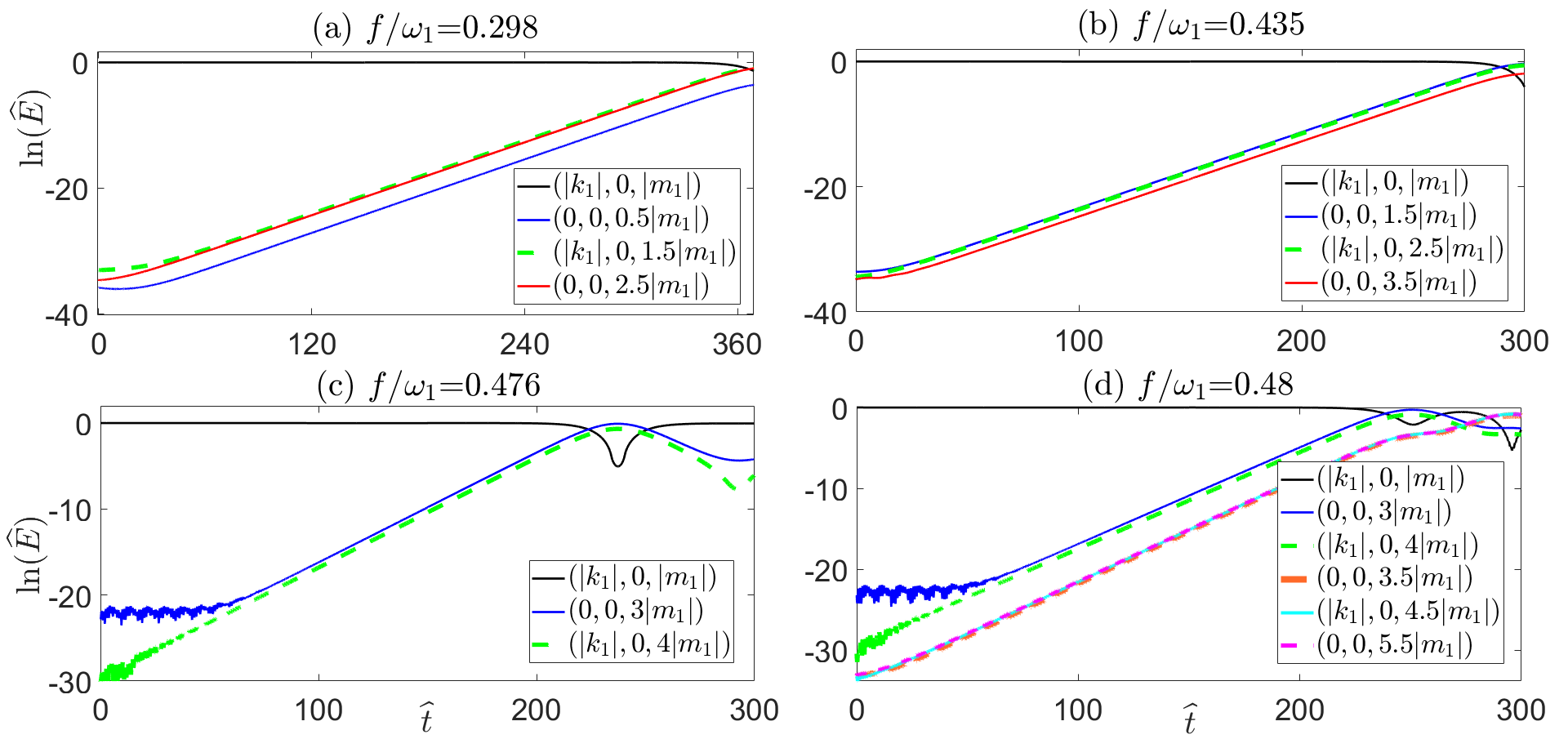}}
 \caption{ Four different 5-wave interactions for primary waves with wavevectors $\mathbf{k}_1 =  (k_1,0,m_1)$ and $\mathbf{k}_5 =  (-k_1,0,m_1)$. }
  \label{fig:k_minus_k_log_energy}
\end{figure} 

\begin{figure}
 \centering{\includegraphics[width=1.0\textwidth]{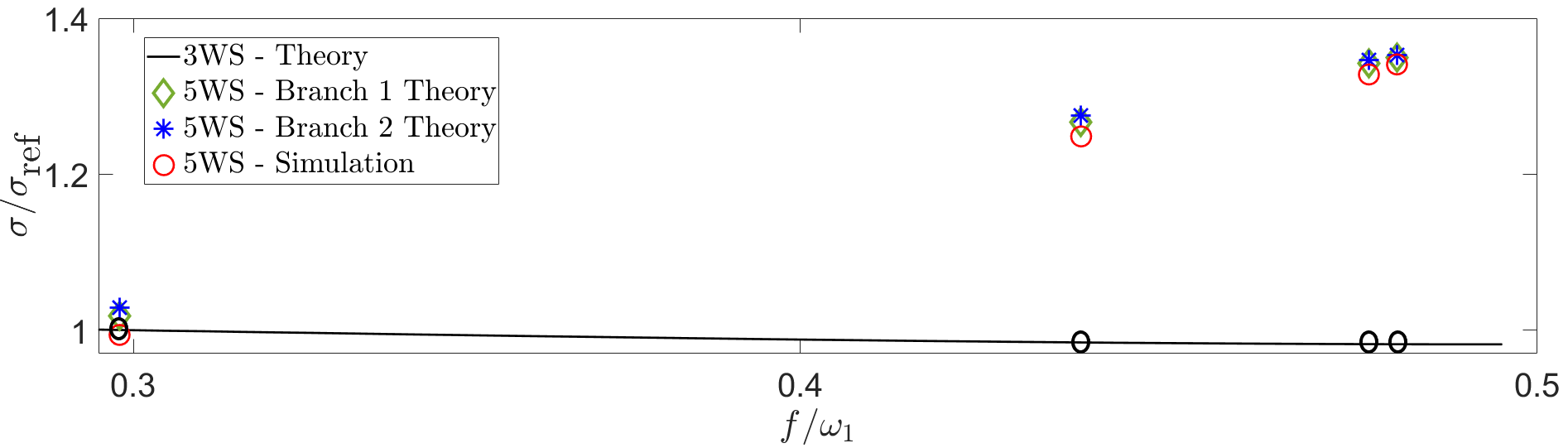}}
 \caption{ Comparison between theoretical growth rates and growth rates obtained from the simulations for $\mathbf{k}_1 =  (k_1,0,m_1)$ and $\mathbf{k}_5 =  (-k_1,0,m_1)$. Red markers indicate results from the simulations. Blue and green markers are predictions from the reduced order model. The black curve plots the variation of the maximum growth rate of all 3-wave systems with $f$.  }
  \label{fig:Simulation_vs_theory_k_minus_k}
\end{figure} 

\subsection{{  { Case $\mathbf{k}_1 =  (k_1,0,m_1)$ and $\mathbf{k}_5 =  (-k_1,0,m_1)$}  }} \label{sec_4_2}

We now validate 5-wave interactions for primary waves propagating in horizontally opposite directions. In this regard, we focus on latitudes where the secondary waves' vertical wavenumbers are multiples of $m_1/2$. Figure \ref{fig:k_minus_k_log_energy} shows the growth of secondary waves for four different $f/\omega_1$ values.  
Figures \ref{fig:k_minus_k_log_energy}(a)--\ref{fig:k_minus_k_log_energy}(b) show energy in three wavevectors increasing exponentially. The three wavevectors encompass both branch-1 and branch-2 secondary waves' wavevectors, and the secondary waves' wavevectors in the simulations are in line with theoretical predictions (theoretical results/wavevectors are not shown in the paper).
The green curve (the wave with non-zero horizontal wavenumber) contains the energy of both leftward and rightward propagating waves. The growth rates estimated from the simulations are much higher than what is expected for a 3-wave interaction. For example, at $f=0.298\omega_1$, the growth rate of the secondary waves is about $30\%$ more than the growth rate of the individual 3-wave interactions that combine to form the 5-wave interaction. Figure \ref{fig:k_minus_k_log_energy}(c) shows only two secondary waves, which are part of the Branch-2 5-wave system. In this case, Branch-1 did not have a growth comparable to Branch-2. Finally, \ref{fig:k_minus_k_log_energy}(d) has three distinct 5-wave systems,
\begin{itemize}
    \item\, System-1 (secondary waves): $(k_1,0,4m_1)$, $(-k_1,0,4m_1)$, $(0,0,-3m_1)$,
    \item\, System-2 (secondary waves): $(k_1,0,4.5m_1)$, $(-k_1,0,4.5m_1)$, $(0,0,-3.5m_1)$,
    \item\, System-3 (secondary waves): $(k_1,0,-4.5m_1)$, $(-k_1,0,-4.5m_1)$, $(0,0,5.5m_1)$.
    \end{itemize}
System-1 is present for both $f/\omega_1=0.476$ and $0.48$ because the change in $f$ is not that significant, and hence the specific interaction is not expected to be detuned significantly. As a result, the system has an exponential growth. Moreover, the growth rates of the waves in any particular system do not vary significantly from the mean value (we observe the maximum deviation from the mean/average growth rate value to be less than $1\%$ of the mean for all the systems).  The growth rates of the waves in System-1 are approximately 0.93 times that of System-2 and System-3 waves. However, System-1 drains the largest amount of energy from the primary waves because the secondary waves in this system have a slightly higher energy at $t=0$.
Growth rates obtained from the reduced order models are once again compared with the growth rates obtained from the numerical simulations, see figure \ref{fig:Simulation_vs_theory_k_minus_k}. When there are multiple branches growing,  the average growth rate of the (two) branches is taken since both branches have nearly the same growth rate. For $f/\omega_1=0.48$ in figure \ref{fig:k_minus_k_log_energy}(d), the average of system-2 and system-3's growth rates is compared with the theoretical growth rate since these are the two resonant Branch-1 and Branch-2 systems at $f/\omega_1=0.48$. It can be seen that theoretical predictions match reasonably well with the simulations. Moreover, similar to \S \ref{sec_4_1}, the growth rates of 5-wave systems are well above the maximum growth of 3-wave systems (shown by the black curve in figure \ref{fig:Simulation_vs_theory_k_minus_k}) for $f/\omega_1>0.4$.   Similar to vertically bounded cases, primary waves with identical vertical wave
numbers and opposite horizontal wavenumber resemble an internal wave mode in a horizontally bounded domain. In this section, we only focused on vertically bounded internal wave modes because they are quite common in the ocean, where the vertical bounds are the air-sea interface
and the ocean topography. Moreover, wave-wave interactions of low mode internal waves are studied extensively to understand their decay in the oceans.  {Horizontally bounded modes can occur in the oceans (for example, in canyons), but We did not give significant attention to them because they are not as common as vertically bounded modes.} We also believe the theoretical results will match the simulations for the horizontally bounded mode, as they did for the vertically bounded mode.

\subsection{Simulations and analysis for $f\approx \omega_1/2$}

In \S \ref{sec_4_1}, we saw that the theoretical growth rates of 5-wave systems are not accurate for $f \approx \omega_1/2$. To test whether the 5-wave systems' growth rate holds near the critical latitude for $\mathbf{k}_1 =  (k_1,0,m_1)$ and $\mathbf{k}_5 =  (-k_1,0,m_1)$, we conducted simulations for three different $f/\omega_1$ values: $f/\omega_1 = 0.496, 0.498$ and $0.499$. 
Moreover, for each $f$, we ran three simulations: one with $\nu = 10^{-6}$  $\textnormal{m}^2\textnormal{s}^{-1}$, one with $\nu = 0.25 \times 10^{-6}$  $\textnormal{m}^2\textnormal{s}^{-1}$, and finally one simulation with hyperviscous terms instead of viscous terms (i.e. by setting $\nu=0$). The hyperviscous operator $-\nu_{H}( {\nabla^2})^4$ is added to the right hand side of  \eqref{eqn:total_mom_equation}--\eqref{eqn:boyancy_equation} with  $\nu_H = 0.25 \times 10^{-6}$ $ \textnormal{m}^8\textnormal{s}^{-1}$. Hyperviscous terms are intended to make the simulation nearly inviscid, and they have been used previously to study PSI \citep{Haze_2011}. All simulations are run for $150$ time periods of the primary wave.  The simulations are stopped before the small-scale secondary waves attain energy comparable to the primary waves. The small-scale waves will break in such cases, and the ensuing turbulence is not resolved and is also not the focus of this study. We are only interested in the growth rate of the secondary waves. 

\begin{figure}
 \centering{\includegraphics[width=1.0\textwidth]{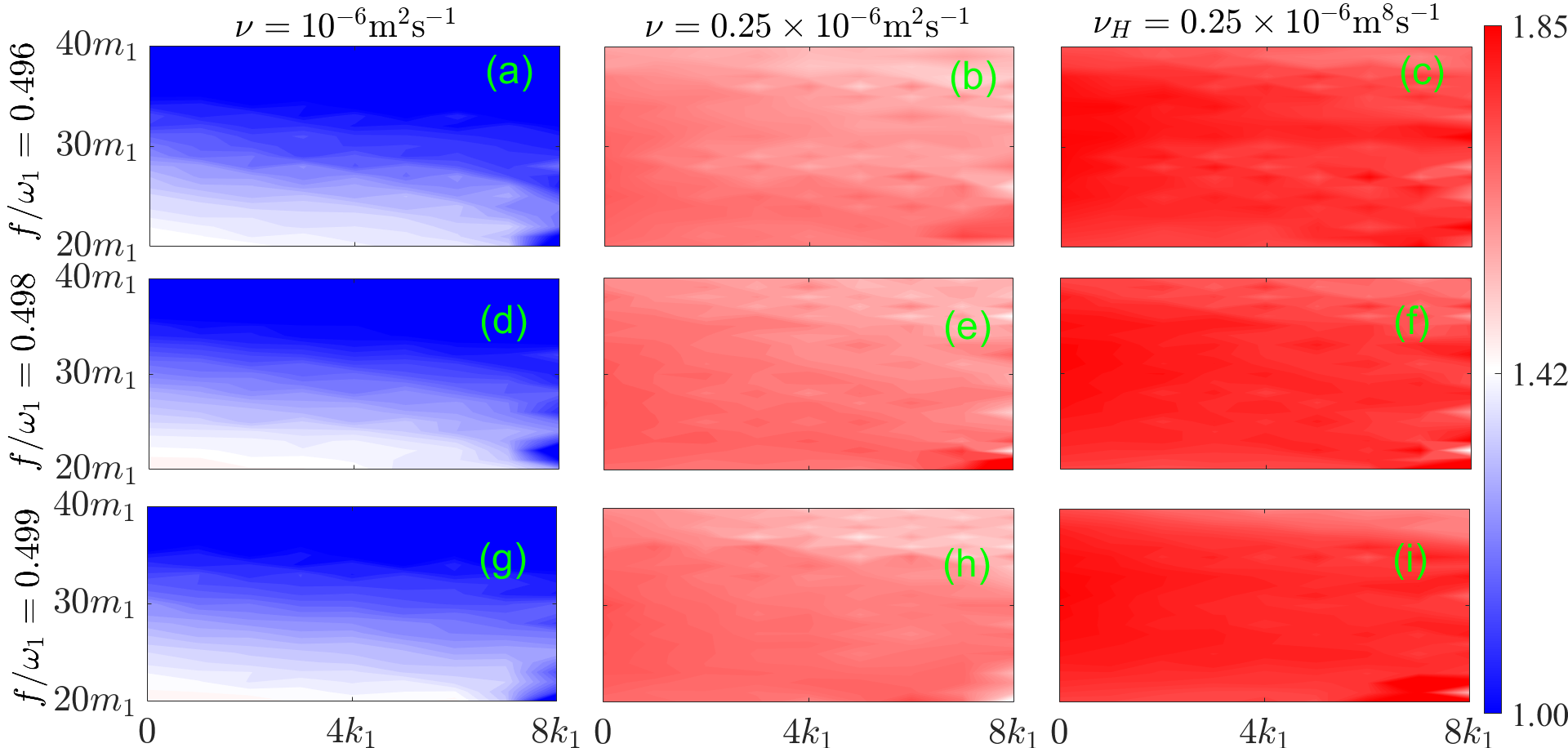}}
 \caption{Growth rate  {colormaps} ($\sigma/\sigma_{\textnormal{cl}}$)   {in the $(k,m)$ plane} for the primary waves with wavevectors $\mathbf{k}_1 =  (k_1,0,m_1)$ and $\mathbf{k}_5 =  (-k_1,0,m_1)$ near the critical latitude ($f/\omega_1 \approx 0.5$). $f/\omega_1=0.496$ for Row-1 ((a), (b) and (c)), $f/\omega_1=0.498$ for Row-2 ((d), (e) and (f)), and $f/\omega_1=0.499$ for Row-3 ((g), (h) and (i)). Viscosity/hyperviscosity values used are as follows: $\nu = 10^{-6} \textnormal{m}^2\textnormal{s}^{-1}$ for Column-1 ((a), (d) and (g)), $\nu = 0.25 \times 10^{-6} \textnormal{m}^2\textnormal{s}^{-1}$ for Column-2 ((b), (e) and (h)), and $\nu_H = 0.25 \times 10^{-6} \textnormal{m}^8\textnormal{s}^{-1}$  for Column-3 ((c), (f) and (i)).
 }
  \label{fig:GR_contours_Simulations}
\end{figure}

Figure \ref{fig:GR_contours_Simulations} shows the non-dimensionalised growth rates ($\sigma/\sigma_{\textnormal{cl}}$) of the secondary waves   in the $(k,m)$ plane for all nine cases. In figure \ref{fig:GR_contours_Simulations}, each row is for a different $f$ value, and for each column, $\nu$ or $\nu_H$ is constant. For the hyperviscous simulations and simulations with the lower viscosity, it can be seen that the non-dimensionalised growth rates are above $\sqrt{2}$ for all three $f$-values (second and third column of figure \ref{fig:GR_contours_Simulations}). Secondary waves with $m=20-40m_1$ have $\sigma/\sigma_{\textnormal{cl}} \approx 1.85$ in the simulations with hyperviscous terms. For all $f$, simulations with $\nu=10^{-6}\textnormal{m}^2\textnormal{s}^{-1}$ have considerably lower growth rates (especially for higher wavenumbers) compared to the other simulations because of the viscous effects. 

We provide the reason for $\sigma/\sigma_{\textnormal{cl}}$ being well above $\sqrt{2}$ using the reduced order model. The dispersion relation for the secondary waves can be rewritten as
\begin{equation}
    (f+\delta\omega)^2 = \frac{N^2(nk_1)^2 + f^2m^2}{(nk_1)^2+m^2},
    \label{eqn:del_omega}
\end{equation}
where $\delta \omega$ is the difference between the wave's frequency ($f+\delta\omega$) and the inertial frequency ($f$), and $n$ is some constant {(but for our purposes, will  primarily be an integer)}.  {$n k_1$ and $m$ are the horizontal and the vertical wavenumber of the secondary waves.}
 {Here, $k_1$ is the zonal wavenumber of the primary waves, but $m$ is \emph{not} the vertical wavenumber of primary waves.}  Near the critical latitude, in a wave-wave interaction, any secondary wave's frequency would be approximately equal to the inertial frequency, implying $\delta \omega\!\ll\! f$. Hence \eqref{eqn:del_omega} leads to
\begin{equation}
    \frac{\delta\omega}{f} \approx \frac{(N^2-f^2)(nk_1)^2}{2f^2[(nk_1)^2+m^2]} \ll 1.
    \label{eqn:del_omega_simplification}
\end{equation}
In scenarios where $N^2\gg f^2$, this yields
\begin{equation}
    m^2 \gg \frac{N^2(nk_1)^2}{2f^2}.
    \label{eqn:wavenumber_relation_crit}
\end{equation}

\begin{figure}
 \centering{\includegraphics[width=1.0\textwidth]{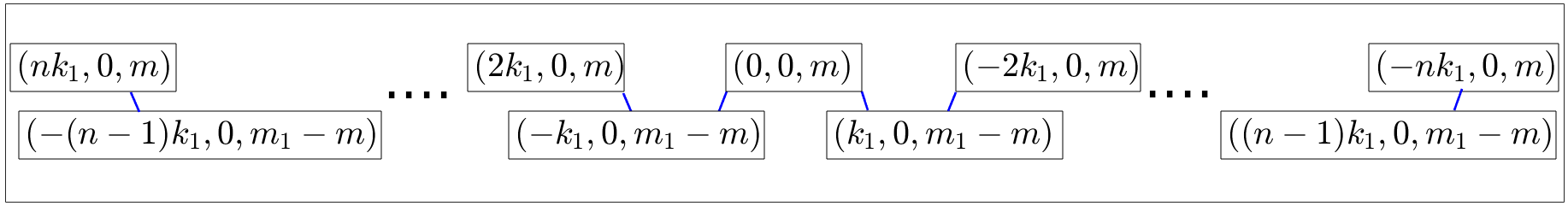}}
 \caption{ A simplified schematic showing how different secondary waves are coupled. Any two wavevectors (boxes) connected by the same blue line can act as a secondary wave combination for the wavevector $\textbf{k}_1 =  (k_1,0,m_1)$ or $\textbf{k}_5 =  (-k_1,0,m_1)$. }
  \label{fig:schematic_coupling}
\end{figure}

\begin{figure}
 \centering{\includegraphics[width=1.0\textwidth]{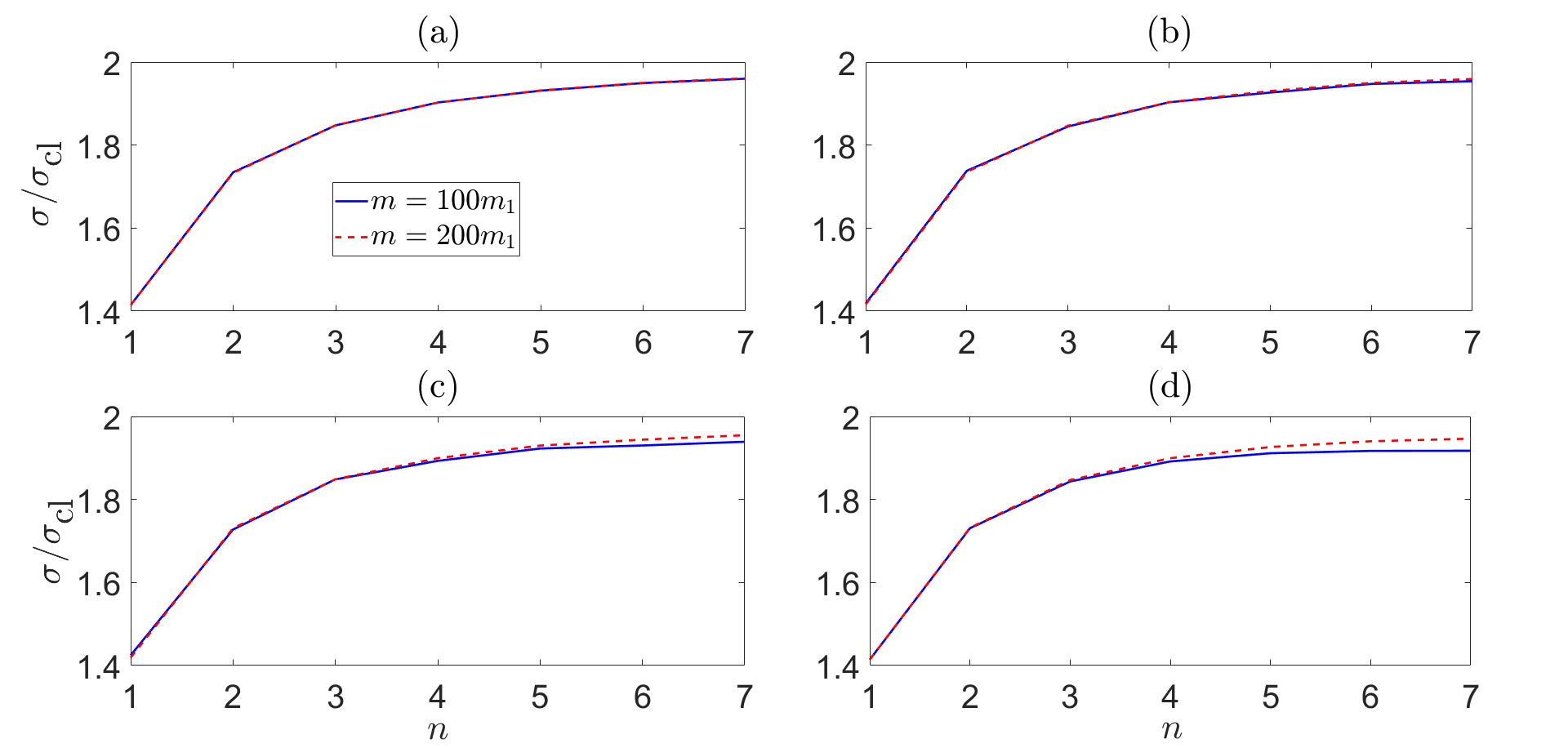}}
 \caption{  Variation of maximum growth rate with $n$ for triad chains near the critical latitude. (a) $\theta=\pi/4$, (b) $\theta=\pi/2$, (c) $\theta=3\pi/4$, and (d) $\theta=\pi$. Two different $m$ values are shown for each $\theta$. }
  \label{fig:f_critical_latitude}
\end{figure}

\noindent Near the critical latitude, $2f \approx \omega_1$. The sum of two secondary waves' frequencies  {would be close to} $\omega_1$ provided their wavenumbers satisfy  \eqref{eqn:wavenumber_relation_crit}.  As a consequence of this special scenario, a chain of coupled triads is possible as shown in figure \ref{fig:schematic_coupling}.  Every box contains the wavevector of a secondary wave. The absolute value of the horizontal wavenumber is lowest at the center of the chain, and it increases in either direction. However, the vertical wavenumber takes only two values. 
Any two boxes that are connected by the same blue line add up to give a primary wave's wavevector. For example, $(2k_1,0,m)+(-k_1,0,m_1-m)$ gives $(k_1,0,m_1)$, which is the wavevector of one of the primary waves. Moreover, $(-k_1,0,m_1-m)+(0,0,m)$ gives $(-k_1,0,m_1)$, which is the other primary wave's wavevector. 
Except for the secondary waves at the ends of the chain, every secondary wave would be forced by both primary waves. Assuming the wavenumbers of the secondary waves in the chain satisfy  \eqref{eqn:wavenumber_relation_crit},  the sum of any two waves' frequencies  {would be close to} $\omega_1$, thus satisfying all the required triad conditions. For a fixed $m$, $\delta \omega$ would increase as $n$ is increased, which is evident from  \eqref{eqn:del_omega}. Hence for very large $n$, the secondary wave's frequency ($f+\delta\omega$) cannot be approximated by $f$ and the sum of two secondary waves' frequencies cannot be approximated by $\omega_1$ simply because $\delta\omega$ would be large. As a result, the triad conditions would not be satisfied for very large $n$. Assuming $\delta \omega$ is negligible up to some $n$,  the wave amplitude equations for the $2n+1$ secondary waves shown in figure \ref{fig:schematic_coupling} can be written in a compact way,
\begin{align}
\dv{ \mathbf{a}}{t} &= \mathcal{H} \mathbf{\bar{a}} \label{eqn:triad_chain_amplitude_equation} \\
\mathbf{a} &= [a_{-n}\hspace{0.2cm} a_{1-n} \hspace{0.2cm} \dots \hspace{0.2cm} a_{n-1} \hspace{0.2cm} a_{n}]^T   \\
\mathcal{H} &= \begin{bmatrix}
 -\mathcal{V}_{-n} & \mathcal{M}_{(-n,1-n)} {A}_1  & 0  & 0  & 0  \\ 
\mathcal{M}_{(1-n,-n)} {A}_1 & -\mathcal{V}_{1-n}  & \mathcal{N}_{(1-n,2-n)} {A}_5  & 0  & 0  \\ 
\vdots & \ddots  & \ddots & \ddots  & \vdots  \\ 
 0  & 0  &\mathcal{M}_{(n-1,n-2)} {A}_1  & -\mathcal{V}_{n-1}  & \mathcal{N}_{(n-1,n)} {A}_5 \\ 
 0 & 0  & 0  & \mathcal{N}_{(n,n-1)} {A}_5   & -\mathcal{V}_{n} \\ 
\end{bmatrix} 
\label{eqn:Matrix_of_connection_of_waves}
\end{align}
where the coefficients $\mathcal{N}_{(i,j)}$ and $\mathcal{M}_{(i,j)}$ are given by
\begin{equation}
    \mathcal{N}_{(i,j)} = \frac{\mathfrak{N}_{(i,5,j)}}{\mathcal{D}_{i}}, \hspace{0.5cm}     \mathcal{M}_{(i,j)} = \frac{\mathfrak{N}_{(i,1,j)}}{\mathcal{D}_{i}}.
\end{equation}
The expression for $\mathfrak{N}_{(i,*,j)}$ is given in Appendix \ref{App:A}.  {Equation \eqref{eqn:triad_chain_amplitude_equation} is an extension of the system given in \eqref{eq:scatter_matrix_scho} to an arbitrary number of secondary waves (using $n=1$ in \eqref{eqn:triad_chain_amplitude_equation} would result in equation \eqref{eq:scatter_matrix_scho}).}
The growth rate for the system given in  \eqref{eqn:triad_chain_amplitude_equation} can be found by calculating the eigenvalues of $\mathcal{H}$. In addition to the $\mathbf{k}_1 =  (k_1,0,m_1)$ and $\mathbf{k}_5 =  (-k_1,0,m_1)$ case, we also analyze the theoretical growth rates for oblique primary waves near the critical latitude using  \eqref{eqn:triad_chain_amplitude_equation}. 
To this end, we consider four $\theta$ values: $\theta=\pi/4$, $\pi/2$, $3\pi/4$, and $\pi$ (see  \eqref{eqn:theta_definition} for the definition of $\theta$). For $\theta \neq \pi$, the primary waves have a non-zero meridional wavenumber ($l_1$). In such cases, the meridional wavenumber of all the secondary waves in the chain is simply assumed to be $l_1/2$. For all four $\theta$ values, figure \ref{fig:f_critical_latitude} shows the gradual increase of the growth rate as $n$ increases for two different $m$ values.  {The $m$ values chosen are $m = 100m_1$ and $m= 200m_1$, where $m_1$ is the primary wave's vertical wavenumber. We chose a high vertical wavenumber for the secondary waves so that condition \eqref{eqn:wavenumber_relation_crit} is satisfied.} For all the $\theta$ values, $\sigma/\sigma_{\textnormal{cl}} \approx 2$ for the higher $n$ values, which is what we observed in the simulation results shown in figure \ref{fig:GR_contours_Simulations}.   { Even though the theoretical value of 2 is obtained for inviscid cases, even for the hyperviscous simulations, we obtained a maximum growth rate of $\approx 1.85$ for secondary waves with vertical wavenumbers in the range $20-40m_1$. This is because the theory is given for vertical wavenumbers $100$ and $200m_1$ to reduce the effects of detuning as much as possible. However, simulating cases that produce secondary waves with very high wavenumbers ($\mathcal{O}(100)m_1$) would be much more computationally expensive. We believe this is one of the main reasons the growth rates are not much closer to 2. Also, simulations with kinematic viscosity $\nu = 10^{-6}$ may not resemble oceanic scenarios completely. This is because even though $\nu = 10^{-6}$ $\textnormal{m}^2\textnormal{s}^{-1}$ is used, the amplitude of the waves is chosen to be low, and this increases the effect of viscosity.}Moreover, for $n=1$, $\sigma/\sigma_{\textnormal{cl}} \approx \sqrt{2}$ which is what we would expect for a 5-wave system with three secondary waves. Interestingly, for an oblique set of primary waves, the results are similar to the 2D case. Hence, 5-wave system growth rates do not apply near the critical latitude for an oblique set of primary waves either. Even though high values of $m$ are used in the reduced order model, simulations show that the resonance can occur even at $m=20-40m_1$. As a result, near the critical latitude, regardless of the $\theta$ value, two primary waves force secondary waves as if they are a single wave with approximately twice the amplitude.

\section{Summary and Conclusions} \label{Section:5}

{In this paper, we use multiple scale analysis to study 5-wave interactions that consist of two plane primary waves with the same frequency but different wavevectors coexisting in a region. We specifically focus on scenarios where both primary waves force a common secondary wave, and this results in a 5-wave system that is composed of two triads. Each of these triads consists of a primary wave and two secondary waves, with one secondary wave common between the two triads. The wavevector and the frequency conditions satisfied by the 5 waves are given in figure \ref{fig:5_wave_triad_schematic}(c).}

For primary waves with wavevectors $(k_1,0,m_1)$ and $(k_1,0,-m_1)$, the 5-wave system is only possible when the common secondary wave's frequency is almost equal to $\omega_1 - f$ (where $\omega_1$ is the primary wave's frequency). The other two secondary waves are near-inertial waves that always propagate in vertically opposite directions. The growth rate of the above-mentioned 5-wave system is higher than the maximum growth rate of 3-wave systems for $f/\omega_1\gtrapprox0.3$. For primary waves with wavevectors $(k_1,0,m_1)$ and $(-k_1,0,m_1)$ (primary waves that propagate in horizontally opposite directions), similar to the previous primary wave combination, the maximum growth rate of 5-wave systems is higher than the maximum growth rate of 3-wave systems for $f/\omega_1\gtrapprox0.3$. For $f/\omega_1\gtrapprox0.3$, the common secondary wave's frequency is nearly equal to $f$ in the most unstable 5-wave system. Moreover, as the common secondary wave's frequency is increased from $f$, the meridional wavenumber increases significantly while the zonal wavenumber of the common secondary wave remains negligible. 

We also study 5-wave systems for cases where the two primary waves are not confined to the same vertical plane. In such scenarios, the dominance of the 5-wave systems increases as the angle between the horizontal wavevectors of the primary waves (denoted by $\theta$) is reduced. Moreover, for any $\theta$, the 5-wave system's instability is dominant over the 3-wave system's instability for $f \gtrapprox 0.3\omega_1$.

\begin{figure}
 \centering{\includegraphics[width=1.0\textwidth]{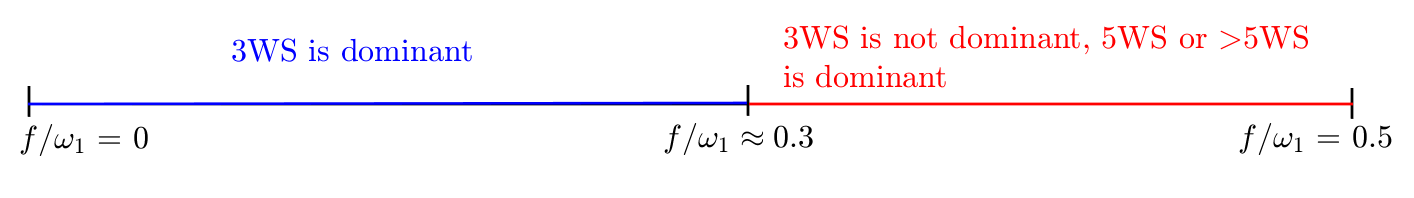}}
 \caption{ A simplified schematic showing which instability mechanism dominates as $f/\omega_1$ is varied. }
  \label{fig:main_result}
\end{figure}

Numerical simulations are conducted to test the theoretical predictions, and the theoretical growth rate of the 5-wave systems matches reasonably well with the results of the numerical simulations for a wide range of $f-$values. However, for all the 2D primary wave combinations considered,  the growth rates from the simulations do not match the theoretical 5-wave systems' growth rate near the critical latitude where $f\approx\omega_1/2$. Near the critical latitude, multiple (more than two) triads become coupled, and they are forced by the two primary waves.  By modifying the reduced order model to account for a chain of secondary waves, the maximum growth rate is shown to be twice the maximum growth rate of all 3-wave systems. Moreover, near the critical latitude, the reduced order model showed similar results for primary waves that are not on the same vertical plane. Hence, near the critical latitude, the 5-wave systems' prediction is not expected to hold for oblique primary waves either. In summary, as shown in figure \ref{fig:main_result}, 5-wave systems will be the dominant instability mechanism for $f/\omega_1 \gtrapprox 0.3$ regardless of the relative orientation of the two primary waves.

We will now provide a brief summary on the applicability of the study to realistic ocean scenarios. In oceans, two or more internal waves can coexist in many scenarios. For example, when an internal wave gets reflected, there would be a finite region where the incident and the reflected wave overlap. Vertically propagating internal waves in the ocean will get reflected when they encounter obstacles in their path, for example, sharp thermocline/pycnocline or the seafloor.  Moreover, tide-topography interactions lead to internal waves with the same frequency but with different wavevectors existing in the same physical space, c.f \citet[figure 7]{sonya_nik_2011} or \citet[figure 9]{richet_2018}.
Interestingly, \citet[figure 10]{richet_2018} shows that the growth rate of near-inertial waves is higher than the maximum growth rate of 3-wave systems at $f/\omega_1 \approx 0.38$. Multiple triad interactions can be a possible explanation for the increased growth rate observed. Our study also highlights the importance of near-inertial waves' role in wave-wave interactions of semidiurnal inertia-gravity waves.

\noindent \textbf{Declaration of interests.} The authors report no conflict of interest. 
   
\noindent \appendix

\noindent \section{Nonlinear coupling coefficients} \label{App:A}

\noindent The quantities $\mathfrak{N}_{(j,p,d)}$ and $\mathfrak{O}_{(j,b,c)}$ are defined so that the nonlinear coefficients can be written in a compact form, and they are given by

\begin{align}
    \mathfrak{N}_{(j,p,d)} = &-{(\omega_p-\omega_d)k_jm_j} \left[ \left( U_p\bar{U}_d(k_j)  + U_p\bar{V}_d l_p - V_p\bar{U}_dl_d + U_p m_p - m_d\bar{U}_d  \right) \right] \nonumber \\
   &- {(\omega_p-\omega_d)l_jm_j} \left[ \left( V_p\bar{V}_d(l_j)  - U_p\bar{V}_d k_d + V_p\bar{U}_d k_p + V_p m_p - m_d\bar{V}_d  \right) \right] \nonumber \\
   &+ {(\omega_p-\omega_d)(l_j^2+k_j^2)} \left[   \bar{V}_d l_p-l_dV_p  + m_j + k_p\bar{U}_d - U_p k_d \right] \nonumber \\
   &+ \textnormal{i}{(l_j^2+k_j^2)} \left[  \bar{U}_d B_p k_p  - U_p\bar{B}_d k_d + \bar{V}_d B_p l_p  - V_p\bar{B}_d l_d  + B_p m_p  - \bar{B}_d m_d      \right] \nonumber \\
   &+ \textnormal{i}{fl_jm_j} \left[ \left( U_p\bar{U}_d(k_j)  + U_p\bar{V}_d l_p - V_p\bar{U}_dl_d + U_p m_p - m_d\bar{U}_d  \right) \right] \nonumber \\
   &- \textnormal{i}{fk_jm_j} \left[ \left( V_p\bar{V}_d(l_j)  - U_p\bar{V}_d k_d + V_p\bar{U}_d k_p + V_p m_p - m_d\bar{V}_d  \right) \right],
   \label{eqn:daughter_NLT}
\end{align}
\begin{align}
     \mathfrak{O}_{(j,b,c)} = &-{(\omega_b+\omega_c)k_jm_j} \left[ \left( U_b{U}_c(k_j)  + U_b{V}_c l_b + V_b{U}_cl_c + U_b m_b + {U}_cm_c  \right) \right] \nonumber \\
   &- {(\omega_b+\omega_c)l_jm_j} \left[ \left( V_b{V}_cl_j  + U_b{V}_c k_c + V_b{U}_c k_b + V_b m_b + m_c{V}_c  \right) \right] \nonumber \\
   &+ {(\omega_b+\omega_c)(l_j^2+k_j^2)} \left[   {V}_c l_b+l_cV_b  + m_j + k_b{U}_c + U_b k_c \right] \nonumber \\
   &+ \textnormal{i}{(l_j^2+k_j^2)} \left[  {U}_c B_b k_b  + U_b{B}_c k_c + {V}_c B_b l_b  + V_b{B}_c l_c  + B_b m_b  + {B}_c m_c      \right] \nonumber \\
   &+ \textnormal{i}{fl_jm_j} \left[ \left( U_b{U}_ck_j  + U_b{V}_c l_b + V_b{U}_cl_c + U_b m_b + m_c{U}_c  \right) \right] \nonumber \\
   &- \textnormal{i}{fk_jm_j} \left[ \left( V_b{V}_cl_j  + U_b{V}_c k_c + V_b{U}_c k_b + V_b m_b + m_c{V}_c  \right) \right], 
   \label{eqn:parent_NLT}
\end{align} 

\noindent where the indices $(j,p,d,b,c)$ are used to denote waves.
Using  \eqref{eqn:daughter_NLT} and \eqref{eqn:parent_NLT}, the nonlinear terms and coefficients used in wave amplitude equations can be written as
\begin{align}
    \mathcal{M}_1 &= \frac{\mathfrak{O}_{(1,2,3)}}{\mathcal{D}_{1}}, \hspace{0.5cm} \mathcal{M}_2 = \frac{\mathfrak{N}_{(2,1,3)}}{\mathcal{D}_{2}}, \hspace{0.5cm} \mathcal{M}_3 = \frac{\mathfrak{N}_{(3,1,2)}}{\mathcal{D}_{3}}, \\
    \mathcal{N}_5 &= \frac{\mathfrak{O}_{(5,3,4)}}{\mathcal{D}_{5}}, \hspace{0.5cm} \mathcal{N}_4= \frac{\mathfrak{N}_{(4,5,3)}}{\mathcal{D}_{4}}, \hspace{0.5cm} \mathcal{N}_3 = \frac{\mathfrak{N}_{(3,5,4)}}{\mathcal{D}_{3}}, \\
        \textnormal{NLT}_1 &= \mathcal{M}_1 \mathcal{D}_{1} a_2 a_3, \hspace{0.5cm}   \textnormal{NLT}_5 = \mathcal{N}_5 \mathcal{D}_{5} a_3 a_4, \\
     \textnormal{NLT}_4 &= \mathcal{N}_4 \mathcal{D}_{4} a_5 \bar{a}_3, \hspace{0.5cm}   \textnormal{NLT}_3 = \mathcal{N}_3 \mathcal{D}_{3} a_5 \bar{a}_4 + \mathcal{M}_3 \mathcal{D}_{3} a_1 \bar{a}_2, \hspace{0.5cm}  \textnormal{NLT}_2 = \mathcal{M}_2 \mathcal{D}_{2} a_1 \bar{a}_3. 
\end{align}

\noindent \section{Derivation of equation \eqref{eqn:growth_rate_inequality}} \label{App:B}

The growth rate of a particular triad with primary wave-1 is given by $\sigma_1 = \sqrt{\bar{\mathcal{M}}_{2}{\mathcal{M}}_{3} |A_1|^2} $. Similarly, the growth rate of a particular triad with primary wave-5 is given by $\sigma_5 = \sqrt{\bar{\mathcal{N}}_{4}{\mathcal{N}}_{3} |A_5|^2} $. If we assume that both primary waves have the same frequency and wavevector norm (which is an assumption that is consistently made in this paper), then it implies that   {the maximum value of  $ \sqrt{\bar{\mathcal{M}}_{2}{\mathcal{M}}_{3} } $ is equal to the maximum value of $\sqrt{\bar{\mathcal{N}}_{4}{\mathcal{N}}_{3}} $.} Mathematically, this can be written as
\begin{equation}
    \textnormal{max}\left(\sqrt{\bar{\mathcal{M}}_{2}{\mathcal{M}}_{3}}\right) =  \textnormal{max}\left(\sqrt{\bar{\mathcal{N}}_{4}{\mathcal{N}}_{3}}\right)  {.}
  \label{eqn:max_cond}  
\end{equation}
For $\nu = 0$ (inviscid flow), the growth rate for a 5-wave system has a simple expression given by
\begin{equation}
        \sigma = \sqrt{\bar{\mathcal{M}}_{2}{\mathcal{M}}_{3} |A_1|^2 + \bar{\mathcal{N}}_{4}{\mathcal{N}}_{3} |A_5|^2}.
        \label{eqn:growth_rate_inviscid_A1}
\end{equation}
Using equation \eqref{eqn:max_cond} in \eqref{eqn:growth_rate_inviscid_A1}, we can arrive at a condition for the upper bound of $\sigma$, which is given by 
\begin{equation}
        \sqrt{\bar{\mathcal{M}}_{2}{\mathcal{M}}_{3} |A_1|^2 + \bar{\mathcal{N}}_{4}{\mathcal{N}}_{3} |A_5|^2} \leq \sqrt{2} \widehat{\sigma}_{1} \hspace{0.5cm} \textnormal{or} \hspace{0.5cm}         \sqrt{\bar{\mathcal{M}}_{2}{\mathcal{M}}_{3} |A_1|^2 + \bar{\mathcal{N}}_{4}{\mathcal{N}}_{3} |A_5|^2} \leq \sqrt{2} \widehat{\sigma}_{5}  {,}
        \label{eqn:growth_rate_inequality_A1}
\end{equation}
where $\widehat{\sigma}_1(\widehat{\sigma}_5)$ is the maximum growth rate of all 3-wave systems of primary wave-1(5).

\bibliography{apssamp}% Produces the bibliography via BibTeX.

\end{document}